\documentclass{article}
\usepackage{geometry}
\usepackage{booktabs} % For formal tables
\usepackage{amssymb}
\usepackage{amsmath}
\usepackage{graphicx}
\usepackage{subfigure}
\usepackage{array}
\usepackage{makecell}
\usepackage{multirow}
\usepackage{diagbox}
\usepackage{xspace}

\newcommand{\np}{\textsc{NP}\xspace}
\newcommand{\nph}{{\np}-hard\xspace}
\newcommand{\npc}{{\np}-complete\xspace}

\newcommand{\myparagraph}[1]{\medskip\noindent\textsf{\bf \small \boldmath #1}}

\newtheorem{problem}{\bf Problem}

\begin{document}

% Page heads
%\markboth{W. Didimo et al.}{Beyond Planarity: Drawing Graphs with Forbidden Crossing Configurations}
%\title[Graph Drawing Beyond Planarity]{Beyond Planarity: Drawing Graphs with Forbidden Crossing Configurations}
\title{A Survey on Graph Drawing Beyond Planarity}

\author{Walter Didimo$^1$,
Giuseppe Liotta$^1$,
Fabrizio Montecchiani$^1$
\\[0.1in]
$^1$Dipartimento di Ingegneria, Universit{\`a} degli Studi di Perugia, Italy\\
\texttt{\small \{walter.didimo,giuseppe.liotta,fabrizio.montecchiani\}@unipg.it}
}

\date{}

\maketitle

\begin{abstract}
Graph Drawing Beyond Planarity is a rapidly growing research area that classifies and studies geometric representations of non-planar graphs in terms of forbidden crossing configurations. Aim of this survey is to describe the main research directions in this area, the most prominent known results, and some of the most challenging open problems.
\end{abstract}

\section{Introduction}\label{se:intro}
In the mid 1980s, the early pioneers of  graph drawing   had the intuition that a drawing with too many edge crossings is harder to read than a drawing of the same graph with fewer edge crossings (see, e.g., ~\cite{DBLP:conf/er/BatiniFN85,DBLP:journals/tse/BatiniNT86,DBLP:journals/tsmc/Carpano80,DBLP:journals/tsmc/SugiyamaTT81}). This intuition was later confirmed by a series of cognitive experimental studies (see, e.g.,~\cite{DBLP:journals/iwc/Purchase00,DBLP:journals/ese/PurchaseCA02,DBLP:journals/ivs/WarePCM02}). As a result, a large part of the existing literature on graph drawing showcases elegant algorithms and sophisticated data structures under the assumption that the input graph is planar, i.e., it admits a drawing without edge crossings. When the input graph is non-planar, crossing minimization heuristics are used to insert a small number of dummy vertices in correspondence of the edge crossings, so to obtain a planarization of the input graph. A crossing-free drawing of the planarization can be computed by using one of the algorithms for planar graphs and then the crossings are reinserted by removing the dummy vertices.
This approach is commonly adopted and works well for graphs of relatively small size, up to a few hundred vertices and edges (see, e.g.,~\cite{dett-gd-99,jm-gds-03}). However, the technological advances of the last twenty years have generated torrents of relational
data that are typically modeled as large graphs with thousand (or more) vertices. These graphs are often hard to visually analyze due, mainly, to their large size which typically implies that a high number of edge crossings is unavoidable even by the most sophisticated planarization approaches. As a consequence, a strong consensus has developed that a new theory of non-planar graph drawing is needed.

In this context, in the early 2000s Mutzel ran an informal experiment with computer scientists at a Dagstuhl workshop, where she presented two different drawings of the same bipartite graph: One has the minimum number of edge crossings, the second has $41\%$ more edge crossings but it has skewness four, which means that all crossings can be removed by deleting four edges in the drawing. All computer scientists found the drawing with skewness four more readable than the one with fewer edge crossings. This is reported in~\cite{DBLP:journals/siamjo/Mutzel01} as anecdotal evidence that the {\em topological properties} of the edge crossings may be more important than their number.

Besides the topological properties  of the edge crossings, the impact of their {\em geometric properties} on the readability of a non-planar drawing was evaluated in a pioneering sequence of user experiments performed in the graph drawing research lab at the University of Sydney. By means of an eye-tracking device, the experiments present statistical evidence that crossings significantly affect human understanding if they form acute angles, but if these angles vary in the range from about $\frac{\pi}{3}$ to $\frac{\pi}{2}$ they guarantee good readability properties~\cite{DBLP:conf/apvis/Huang07,DBLP:conf/apvis/HuangHE08,DBLP:journals/vlc/HuangEH14}.

These empirical experiences suggest that a new theory of non-planar graph drawing can be developed under the assumption that not only the number of edge crossings but also their (topological and/or geometric) properties have an impact on the readability of a diagram. 
%In this respect, it may be worth recalling that planar graphs can be expressed in terms of forbidden subgraphs: A graph is planar if and only if it does not contain a subdivision of $K_5$ or $K_{3,3}$. 
Hence, a natural step towards understanding non-planar representations of graphs is to classify and study them in terms of {\em forbidden crossing configurations}. This is, in a broad sense, the aim and scope of the rapidly growing research area of {\em graph drawing beyond planarity}. Table~\ref{ta:families} reports some examples of beyond-planar graphs with a description of their forbidden crossing configurations\footnote{The table only shows a subset of the graph families described in this paper.}. 

\myparagraph{Overview and paper organization.} The goal of this survey is to classify and describe prominent results and promising research directions in the fertile area of graph drawing beyond planarity. The survey addresses the following questions.

\begin{description}
\item[Q1] What are the forbidden crossing configurations and the main research directions that have been studied so far?
\item[Q2] For each such direction, what are the main combinatorial results and which algorithms have been designed, experimented, and engineered?
\item[Q3] What are the most relevant open problems and the least explored research directions in this area?
\end{description}

In order to answer Question~{\em Q1}, in Section~\ref{se:taxonomy} we define graph families that avoid forbidden crossing configurations and we present a taxonomy of the main research directions in the area of graph drawing beyond planarity. Sections~\ref{se:density}--\ref{se:experiments-engineering} address Question~{\em Q2}, using the taxonomy as a guideline to classify the main combinatorial and algorithmic results. At the end of each section we shall address Question~{\em Q3} by discussing some relevant open problems. Question~{\em Q3} is further discussed in Section~\ref{se:conclusions}. Basic definitions on graph drawing can be found in Section~\ref{se:terminology}. 
%Additional references to papers that cover topics beyond the scope of this survey, which are however related to graph drawing beyond planarity, are in Section~\ref{se:related}. 
%
%
\section{Basic definitions on Graph Drawing}\label{se:terminology}
Let $G=(V,E)$ be a graph. If not otherwise specified, we assume that $G$ may contain multiple edges but no self-loops. A \emph{drawing} $\Gamma$ of $G$ maps each vertex $v \in V$ to a distinct point $p_v$ of the plane and each edge $(u,v) \in E$ to a simple Jordan arc with endpoints $p_u$ and $p_v$. In notation and terminology, we make no distinction between the vertices and edges of a graph and the points and arcs representing them, respectively. Two edges of $\Gamma$ \emph{cross} if they have a point in common distinct from their endpoints; this point is a \emph{crossing}. We assume that an edge does not contain a vertex other than its endpoints, no two edges meet tangentially, and no three edges share a crossing. $\Gamma$ is {\em simple} if any two edges intersect in at most one point, which is either a common endpoint or an interior point where the two edges properly cross. Thus in a simple drawing two adjacent edges do not cross and two non-adjacent edges cross at most once.

\begin{figure}
\centering
\subfigure[]{\includegraphics[width=0.18\columnwidth,page=1]{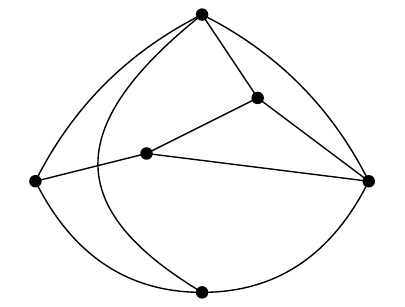}\label{fi:rotsys-1}}\hfil
\subfigure[]{\includegraphics[width=0.18\columnwidth,page=2]{figs/rotation-system}\label{fi:rotsys-2}}\hfil
\subfigure[]{\includegraphics[width=0.18\columnwidth,page=3]{figs/rotation-system}\label{fi:rotsys-3}}\hfil
\subfigure[]{\includegraphics[width=0.18\columnwidth,page=4]{figs/rotation-system}\label{fi:rotsys-4}}
\caption{(a)-(b) Two drawings with the same rotation system but with different embeddings. (c)-(d) Two planar embeddings that differ for the choice of the outer face.\label{fi:rotsys}}
\end{figure}

A drawing $\Gamma$ of $G$ divides the plane into topologically connected regions, called \emph{faces}. The infinite region is called the \emph{external face}; the other regions are the \emph{internal faces}. Note that the  boundary of a face may contain both vertices of the graph and crossings. An \emph{embedding} of $G$ is an equivalence class of drawings of $G$ under homeomorphism of the plane, i.e., is a class of drawings of $G$ that define the same set of (external and internal) faces. A graph with a fixed embedding is called an \emph{embedded graph}. A drawing without crossings is \emph{planar}. A \emph{planar graph} is a graph that admits a planar drawing. A \emph{planar embedding} is the embedding of a planar drawing. A planar graph with a fixed planar embedding is an \emph{embedded planar graph}, or briefly a \emph{plane graph}. Note that, an embedding of a graph $G$ defines, for each vertex $v$ and for each crossing $c$, the clockwise circular order of the edges incident to $v$ and to $c$. A \emph{rotation system} of $G$ is a relaxation of an embedding of $G$ in which the clockwise circular order of the edges is fixed only for each vertex, while no information is given about which pairs of edges cross and in which order. If $G$ is planar, a planar embedding of $G$ corresponds to a rotation system plus the choice of the external face (see Fig.~\ref{fi:rotsys}).

In a \emph{polyline drawing} of a graph each edge is a chain of segments; a \emph{bend} is a point where two segments of the same edge meet. A \emph{$k$-bend drawing} is a polyline drawing with at most $k$ bends per edge. A $0$-bend drawing is also called a \emph{straight-line drawing}. 

We remark that, in the literature, a drawing of a graph is sometimes called a \emph{topological graph} and a straight-line drawing is sometimes called a \emph{geometric graph}. In the remainder of this paper we shall use the pair of terms drawing and topological graph, and the pair straight-line drawing and geometric graph,  interchangeably.
\section{Forbidden Configurations and Main Research Directions}\label{se:taxonomy}
In Section~\ref{ss:families} we survey forbidden crossing configurations; in Section~\ref{ss:taxo} we present a taxonomy of the most explored research directions in graph drawing beyond planarity.

\subsection{Graph Families}\label{ss:families}
The beyond-planar graph families considered in this survey are defined as graphs that admit drawings not containing specific \emph{forbidden configurations}, that is, sets of edges that violate some desired topological or geometric property of the edge crossings. Table~\ref{ta:families} provides a schematic illustration of some beyond-planar graph families, among the most studied in the literature. For each family $X$, we define what a drawing of type $X$ is. A graph belongs to family $X$ if it admits a drawing of type $X$. Definitions based on first-order logic formulas have been recently proposed for many of these types of drawings~\cite{DBLP:journals/jgaa/Brandenburg17}.

\myparagraph{$k$-planar drawings.} A \emph{$k$-planar} drawing $(k \geq 1)$ does not contain an edge crossed more than $k$ times. The family of $k$-planar graphs, for $k=1$, was introduced in 1965 in the context of the simultaneous vertex-face coloring of planar graphs~\cite{R65}. The study of $k$-planar graphs, for $k > 1$, was proposed for the first time as a tool for finding lower bounds on the crossing number of a graph~\cite{DBLP:journals/combinatorica/PachT97}, i.e., on the minimum number of crossings in a drawing of a graph.

\myparagraph{$k$-quasi planar drawings.} A \emph{$k$-quasi planar} drawing $(k \geq 3)$ does not have $k$ mutually crossing edges. The first results about $k$-quasi planar graphs date back in the 80s and 90s, when the problem of determining the maximum number of edges for these graphs was addressed~\cite{DBLP:journals/dcg/AlonE89,DBLP:journals/dcg/PachT94}, answering questions posed even earlier~\cite{avitalhanani66,kupitz1979,akiyamaalon89}.

\begin{table}[p]
	\centering
	\caption{Examples of beyond-planar graph families.\label{ta:families}}\smallskip{
		\renewcommand{\arraystretch}{0.9}
		\begin{tabular}{|>{\centering\arraybackslash}m{1in}|>{\centering\arraybackslash}m{1in}||>{\centering\arraybackslash}m{1.8in}|>{\centering\arraybackslash}m{1in}|}
			\hline
			\multicolumn{2}{|c||}{\textsc{\small Graph Family}} & \multicolumn{2}{|c|}{\textsc{\small Forbidden Configuration}}\\
			\hline
			\textsc{\small Name} &  \textsc{\small Example} & \textsc{\small Description} & \textsc{\small Illustration}\\
			\hline
			{\small $k$-planar ($k \ge 1$)} &
			\setlength\fboxrule{0pt}
			\fbox{\includegraphics[scale=0.4,page=1]{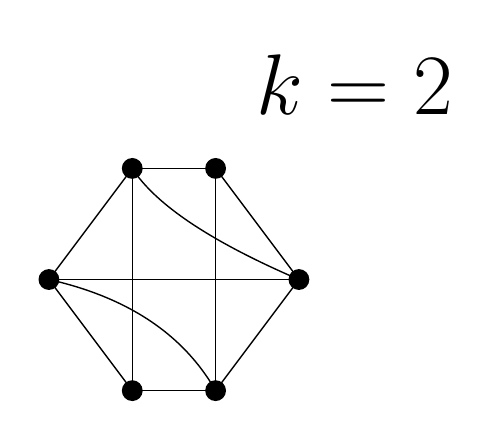}}
			& {\small An edge crossed more than $k$ times} &
			\setlength\fboxrule{0pt}
			\fbox{\includegraphics[scale=0.4,page=1]{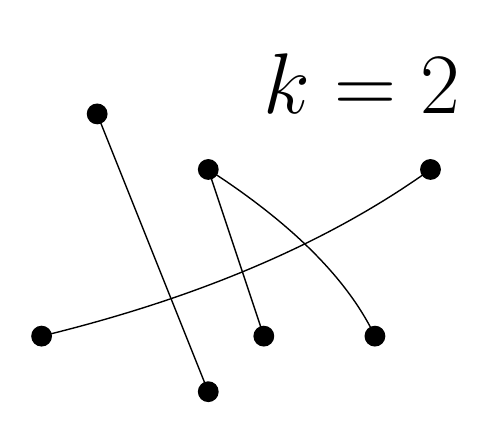}}
			
			\\\hline
			{\small $k$-quasi planar ($k \ge 3$)}&
			\setlength\fboxrule{0pt}
			\fbox{\includegraphics[scale=0.4,page=2]{figs/good-examples}}
			& {\small $k$ pairwise crossing edges} &
			\setlength\fboxrule{0pt}
			\fbox{\includegraphics[scale=0.4,page=2]{figs/forbidden-configurations}}
			
			\\\hline
			{\small skewness-$k$ $(k \geq 1)$} &
			\setlength\fboxrule{0pt}
			\fbox{\includegraphics[scale=0.4,page=7]{figs/good-examples}}
			& {\small Set of crossings not covered by at most $k$ edges} &
			\setlength\fboxrule{0pt}
			\fbox{\includegraphics[scale=0.4,page=7]{figs/forbidden-configurations}}
			
			\\\hline
			%$k$-bend ACE$_\alpha$ ($k \ge 0$) ($0 < \alpha < \frac{\pi}{2}$)& & & \\\hline
			%$k$-bend ACL$_\alpha$ ($k \ge 0$) ($0 < \alpha < \frac{\pi}{2}$)& & & \\\hline
			{\small fan-planar} &
			\setlength\fboxrule{0pt}
			\fbox{\includegraphics[scale=0.4,page=4]{figs/good-examples}}
			& {\small Two independent edges that cross a third one or two adjacent edges that cross another edge from different ``sides''} &
			\setlength\fboxrule{0pt}
			\fbox{\includegraphics[scale=0.4,page=4]{figs/forbidden-configurations}}
			
			\\\hline
			{\small $2$-fan-crossing-free (fan-crossing-free)} &
			\setlength\fboxrule{0pt}
			\fbox{\includegraphics[scale=0.4,page=5]{figs/good-examples}}
			& {\small Two adjacent edges that cross a third one} &
			\setlength\fboxrule{0pt}
			\fbox{\includegraphics[scale=0.4,page=5]{figs/forbidden-configurations}}
			
			\\\hline
			{\small straight-line RAC} &
			\setlength\fboxrule{0pt}
			\fbox{\includegraphics[scale=0.4,page=3]{figs/good-examples}}
			& {\small Two edges crossing at an angle smaller than $\frac{\pi}{2}$}  &
			\setlength\fboxrule{0pt}
			\fbox{\includegraphics[scale=0.4,page=3]{figs/forbidden-configurations}}
			
			\\\hline
			{\small $\alpha$-SHPED ($0 < \alpha < \frac{1}{2}$)} &
			\setlength\fboxrule{0pt}
			\fbox{\includegraphics[scale=0.4,page=6]{figs/good-examples}}
			& {\small Two $\alpha$-stubs that cross} &
			\setlength\fboxrule{0pt}
			\fbox{\includegraphics[scale=0.4,page=6]{figs/forbidden-configurations}}
			
			\\
			\hline
		\end{tabular}
	}
\end{table}

\myparagraph{Skewness-$k$ drawings.} A \emph{skewness-$k$} drawing $(k \geq 1)$ is such that the removal of at most $k$ edges makes the drawing planar. In terms of forbidden configuration it means that the drawing does not contain a set of crossings not covered by (at most) $k$ edges. A graph has  \emph{skewness} $k$ if it admits a skewness-$k$ drawing. Graphs with skewness $k$ are mainly studied for $k=1$, under the name of \emph{almost planar} (or \emph{near planar}) graphs, especially in terms of crossing number~\cite{DBLP:journals/algorithmica/CabelloM11,DBLP:journals/siamcomp/CabelloM13}. The problem of efficiently computing a skewness-$k$ drawing of a graph $G=(V,E)$ ($k \geq 1$) with the minimum number of crossings and with a fixed planar subgraph $G'=(V,E \setminus F)$, where $|F|=k$, was also studied~\cite{DBLP:journals/algorithmica/GutwengerMW05,DBLP:conf/compgeom/ChimaniH16}. For $k=1$, this problem is linear-time solvable and the solution gives an approximation to the crossing number of the almost planar graph $G$~\cite{DBLP:journals/algorithmica/GutwengerMW05}.

\myparagraph{$k$-apex drawings.}
A \emph{$k$-apex} drawing $(k \geq 1)$ is such that the removal of at most $k$ vertices makes the drawing planar. It means that the drawing does not contain a set of crossing edges not covered by (at most) $k$ vertices. It is immediate to see that a skewness-$k$ drawing is also a $k$-apex drawing, but not vice versa.
It is known that $1$-apex graphs, mainly recognized as \emph{apex graphs} in the literature, are closed under the operation of taking minors. They have connections with other aspects of graph minor theory (see, e.g.,~\cite{DBLP:journals/combinatorica/RobertsonST93,DBLP:books/daglib/0076841}) and play a role in the relations between treewidth and graph diameter~\cite{DBLP:journals/algorithmica/Eppstein00,DBLP:journals/algorithmica/DemaineH04}. The problem of efficiently computing an apex drawing with minimum number of crossings and an identified apex vertex was studied~\cite{DBLP:conf/soda/ChimaniGMW09}.

\myparagraph{$(k,l)$-grid-free drawings.} For $k,l \geq 1$, a \emph{$(k,l)$-grid-free} drawing does not contain a \emph{$(k,l)$-grid}, i.e., two groups of $k$ and $l$ edges, respectively, such that each edge of the first group crosses all the edges of the second group~\cite{DBLP:journals/gc/PachPST05}. If the $k$ edges are incident to the same vertex, the $(k,l)$-grid is \emph{radial}; if the $l$ edges are also incident to the same vertex, the $(k,l)$-grid is \emph{biradial}. A $(k,l)$-grid is \emph{natural} if all its edges are independent and no two edges in the same group cross. 
%A $(k,k)$-grid is also called a \emph{$k$-grid}.

\myparagraph{$k$-fan-crossing-free drawings.} A \emph{$k$-fan-crossing-free} drawing does not contain $k \ge 2$ adjacent edges that cross a third one. A $2$-fan-crossing-free drawing is also called \emph{fan-crossing-free} ~\cite{DBLP:journals/algorithmica/CheongHKK15}. The class of $k$-fan-crossing-free graphs coincides with  the class of radial $(k,1)$-grid-free graphs, for $k \geq 2$.

\begin{figure}[tb]
\centering
\subfigure[]{\includegraphics[width=0.24\columnwidth,page=1]{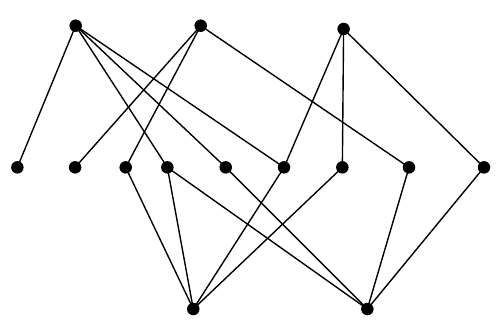}\label{fi:fan-planarity-1}}\hfill
\subfigure[]{\includegraphics[width=0.24\columnwidth,page=2]{figs/fan-planarity}\label{fi:fan-planarity-2}}\hfill
\subfigure[]{\includegraphics[width=0.24\columnwidth,page=3]{figs/fan-planarity}\label{fi:fan-planarity-3}}\hfil
\subfigure[]{\includegraphics[width=0.24\columnwidth,page=4]{figs/fan-planarity}\label{fi:fan-planarity-4}}
\caption{(a) A fan-planar drawing of a graph $G$ with 12 crossings. (b) A confluent drawing of $G$ with 3 crossings. (c) A fan-planar drawing with 16 crossings of another graph $G'$. (d) A confluent drawing of $G'$ with 8 crossings.\label{fi:fan-planarity}}
\end{figure}

\myparagraph{Fan-planar drawings.} A \emph{fan-planar} drawing does not contain two independent edges that cross a third one or two adjacent edges that cross another edge from different ``sides''~\cite{DBLP:journals/corr/KaufmannU14}. This family of drawings is somehow opposite to the class of fan-crossing-free drawings. From a practical point of view, fan-planar drawings can be used to create drawings with few edge crossings per edge in a \emph{confluent drawing} style~\cite{DBLP:journals/jgaa/DickersonEGM05} (see, e.g., Fig.~\ref{fi:fan-planarity}).
Note that a fan-planar drawing cannot contain a $(k,l)$-grid that is not radial, for $k \geq 2$.

\myparagraph{RAC drawings.} A straight-line drawing such that any two crossing edges form $\frac{\pi}{2}$ angles at their crossing point is \emph{straight-line RAC} (RAC stands for Right Angle Crossing). The introduction of RAC drawings~\cite{DBLP:journals/tcs/DidimoEL11} was motivated by cognitive studies that suggest a positive correlation between large crossing angles and human understanding of graph visualizations~\cite{DBLP:conf/apvis/Huang07,DBLP:conf/apvis/HuangHE08,DBLP:journals/vlc/HuangEH14}, and by the common use of large angle crossings in handmade real-world diagrams, such as metro maps~\cite{vignelli2008,roberts2012}. Clearly, a straight-line RAC drawing is fan-crossing-free. RAC drawings with edge bends are also studied in the literature and will be discussed in this survey.

\myparagraph{ACE$\alpha$ and ACL$\alpha$ drawings.} \emph{ACE$\alpha$} and \emph{ACL$\alpha$} drawings are variants of RAC drawings that are parametric in a value $\alpha \in (0,\frac{\pi}{2})$. In a straight-line ACE$\alpha$ drawing any two crossing edges form an angle \emph{equal to} $\alpha$. Graphs that admit this type of drawing were originally introduced with the name of $\alpha$AC$^=$ graphs~\cite{aft-sgapd-10,DBLP:journals/siamdm/AckermanFT12}. In a straight-line ACL$\alpha$ drawing the value of any crossing angle is \emph{at least} $\alpha$. These drawings were independently introduced by two different works; in~\cite{DBLP:journals/mst/GiacomoDLM11} they are named LAC$_\alpha$ drawings and in~\cite{DBLP:journals/cjtcs/DujmovicGMW11} they are called $\alpha$AC drawings. As for RAC drawings, ACE$\alpha$ and ACL$\alpha$ drawings with edge bends are also considered in this survey.

\myparagraph{Partial edge drawings ($\alpha$-SHPEDs).} In an \emph{$\alpha$-SHPED} (Symmetric Homogeneous Partial Edge Drawing) $(0<\alpha<\frac{1}{2})$, each edge $(u,v)$ is (partially) drawn as a pair of straight-line segments, called \emph{$\alpha$-stubs}, one incident to $u$, the other incident to $v$, and each being a fraction $\alpha$ of segment $\overline{uv}$; the two stubs do not cross. The idea behind this drawing style was introduced to visualize network overload between switches of the AT\&T long distance telephone network in the U.S.~\cite{bew-vnd-TVCG95}. The $\alpha$-SHPED model has been formally defined only recently~\cite{DBLP:conf/fun/BruckdorferK12}, and has inspired both theoretical~\cite{JGAA-438} and practical research (see, e.g.,~\cite{DBLP:conf/iisa/BinucciLMT16,DBLP:conf/iisa/Bruckdorfer0L15}). An extension of this model for graphs with maximum degree four where the drawings are orthogonal and have at most one bend per edge is also studied~\cite{DBLP:journals/jgaa/BruckdorferKM14}.

\myparagraph{$k$-gap-planar drawings.}  In a \emph{$k$-gap-planar drawing} it is possible to map each crossing  to one of the two corresponding crossing edges so that, for each edge $e$ at most $k$ crossings are mapped to $e$.
This family generalizes $k$-planar graphs and was introduced in~\cite{DBLP:conf/gd/Bae17} with a practical motivation inspired by edge casing, a method commonly used to alleviate the visual clutter caused by crossing lines~\cite{DBLP:conf/siggraph/AppelRS79,DBLP:journals/comgeo/EppsteinKMS09}. In a cased drawing of a graph, each crossing is resolved by locally interrupting one of the two crossing edges, and a $k$-gap-planar graph can be equivalently defined as a graph that admits a cased drawing in which each edge has at most $k$ gaps.

\myparagraph{$H$-self-intersecting-free drawings.} Given a graph $H$, a straight-line \emph{$H$-self-intersecting-free} drawing (\emph{$H$-sif} drawing) does not contain a self-intersecting geometric graph isomorphic to $H$. This specializes the more general definition of \emph{$H$-free} drawing, i.e., a drawing containing no graph isomorphic to $H$. Typical forbidden configurations considers self-intersecting paths or cycles~\cite{DBLP:conf/compgeom/PinchasiR03,DBLP:journals/ejc/PachPTT04}.

\myparagraph{Planarly-connected drawings.} In a \emph{planarly-connected drawing} each pair of crossing edges is independent and there is a crossing-free edge that connects their endpoints~\cite{DBLP:journals/jgaa/Ackerman17}. As it will be discussed in Section~\ref{se:relationships}, this family includes meaningful subfamilies of $1$-planar and fan-planar drawings.

\subsection{Main Research Directions}\label{ss:taxo}

\begin{figure}[tb]
	\centering
	\includegraphics[width=1\columnwidth]{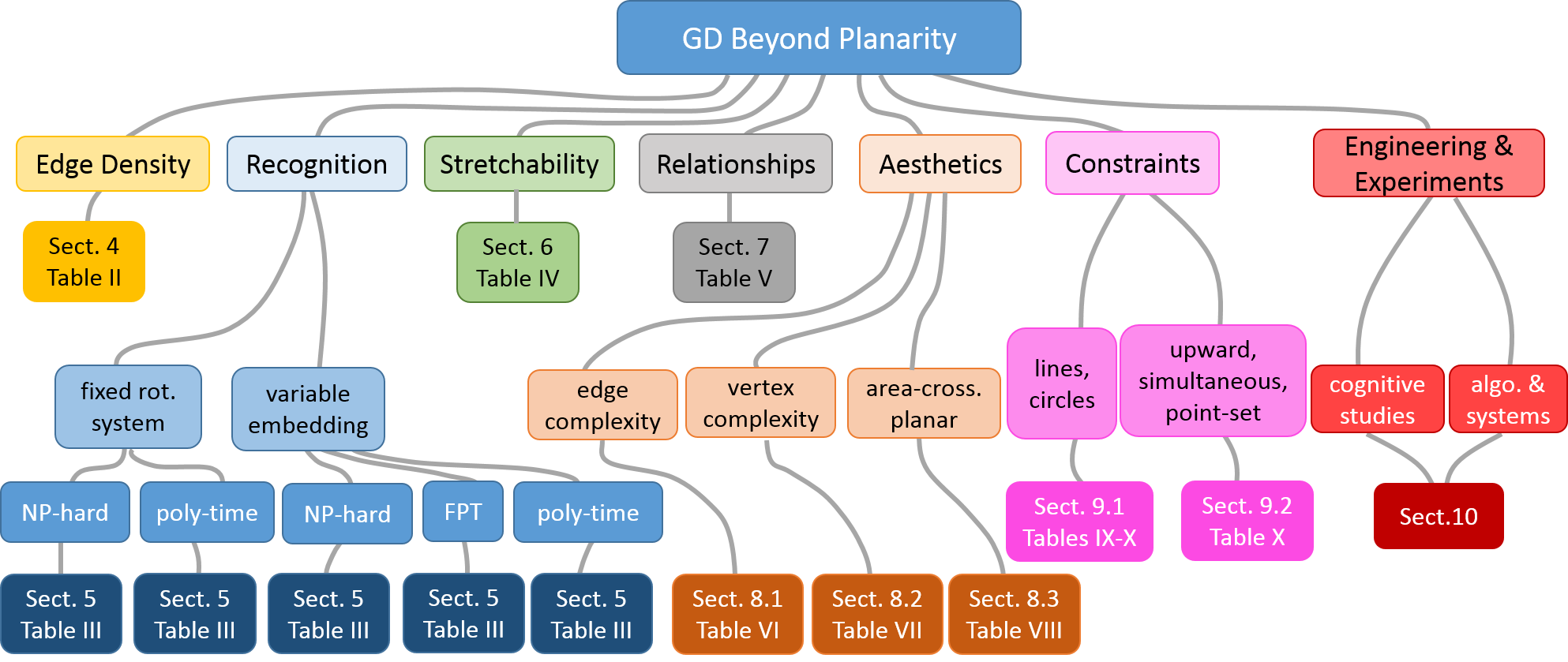}
	\caption{Tree that summarizes the taxonomy used in this survey.}\label{fi:tree}
\end{figure}

The tree of Fig.~\ref{fi:tree} synthetically describes a taxonomy of the graph drawing beyond planarity area. The first-level nodes identify the main research directions that have been studied so far; we classify the results in the literature according to these main research directions. The intermediate nodes are refinements of these main research directions. The leaves refer to the sections of this survey where the research directions are described in detail; most of these sections report one or more summarizing tables, which are also referred in the corresponding leaves. Hereunder we shortly describe the main research directions.

\myparagraph{Density.} Since the mid-1960s different authors (see, e.g.,~\cite{avitalhanani66,b-78,kupitz1979}) initiated the study of the following Tur\'an-type problem: What is the maximum number of edges that a drawing of a graph can have without containing a given type of forbidden crossing configuration? Since then this question has been long studied for all families of beyond-planar drawings. Results about the edge density of beyond-planar graphs are discussed in Section~\ref{se:density}.

\myparagraph{Recognition.} In contrast with planarity testing, which is solvable in linear time~\cite{DBLP:journals/jacm/HopcroftT74}, recognizing whether a graph belongs to a certain family of beyond-planar graphs is computationally hard for most of the cases. For some subfamilies of beyond-planar graphs and/or under additional restriction of the input the recognition problem is polynomial-time solvable. Results about the recognition problem of different families of beyond-planar graphs are surveyed in Section~\ref{se:recognition}.

\myparagraph{Stretchability.} The stretchability problem asks the following question: Let $G$ be a topological graph with a fixed embedding; does $G$ admit a drawing where all edges are straight-line segments and such that the given embedding is preserved? Remarkably, while the stretchability of planar drawings has been studied since the mid 1930s (see, e.g.,~\cite{wagner-36,fary-48,stein-51}), the question has received only little attention when the drawing is non-planar and some crossing configurations are forbidden. Results about the stretchability of beyond-planar graphs are discussed in Section~\ref{se:stretchability}.

\myparagraph{Relationships.} Studying the combinatorial relationships between different families of beyond-planar graphs is a fundamental step towards developing a comprehensive theory of graph drawing beyond planarity. The typical question in this context is the following: Let $F$ and $F'$ be two forbidden types of crossing configurations (for example, $F$ may be ``four mutually crossing edges'' and $F'$ may be ``an edge crossed by three distinct edges''). If a graph $G$ admits a drawing where $F$ is forbidden, does $G$ also admit a drawing where $F'$ is forbidden? Results about relationships of inclusion or of intersection between families of beyond-planar graphs are described in Section~\ref{se:relationships}.

\myparagraph{Aesthetics.} In addition to requiring that some types of edge crossings are forbidden in a non-planar drawing, one can pursue some geometric optimization goals, often called {\em aesthetic requirements}, such as minimizing the drawing area for a given resolution rule, maximizing the drawing aspect ratio, minimizing the number of different slopes used to draw the edges or the number bends per edge. Such aesthetic requirements have a strong impact on the visual complexity of a drawing (see, e.g.,~\cite{dett-gd-99,kw-dg-01}). Results about this research direction are presented in Section~\ref{se:aesthetics}.

\myparagraph{Constraints.} Depending on the type of graph and/or application, additional constraints can be imposed on a drawing. For example, for bipartite graphs a typical constraint is to represent the vertices of each partition set on one of two parallel lines, which is a fundamental step in the so-called ``Sugyiama approach'' for layered drawings~\cite{DBLP:journals/tsmc/SugiyamaTT81}. Other constraints require that all vertices are collinear in the so-called $k$-page drawing model~\cite{Ollmann-73,bk-79} or co-circular in the circular layout model~\cite{DBLP:reference/crc/SixT13}. Results about constrained beyond-planar graphs are surveyed in Section~\ref{se:constraints}.

\myparagraph{Engineering and Experiments.} The road towards an effective technology transfer of the algorithmic solutions developed in the area of graph drawing beyond planarity has just begun. The initial steps in this direction are summarized in Section~\ref{se:experiments-engineering}. 
\section{Edge Density}\label{se:density}
\begin{figure}[tb]
	\centering
	\subfigure[]{\includegraphics[width=0.25\columnwidth,page=1]{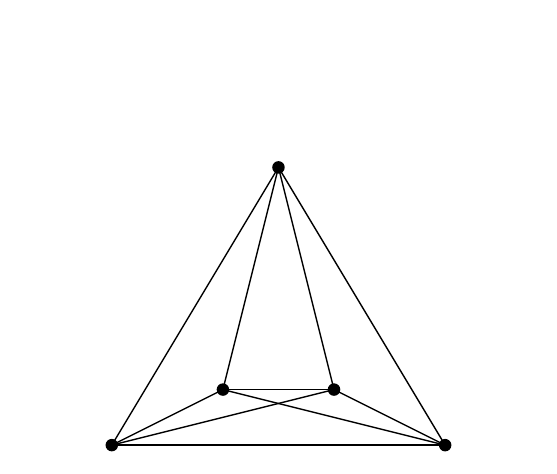}\label{fi:maximally-dense}}\hfill
	\subfigure[]{\includegraphics[width=0.25\columnwidth,page=2]{figs/density-definitions}\label{fi:maximal}}\hfill
	\subfigure[]{\includegraphics[width=0.25\columnwidth,page=3]{figs/density-definitions}\label{fi:optimal}}\hfill
	\subfigure[]{\includegraphics[width=0.25\columnwidth,page=4]{figs/density-definitions}\label{fi:optimal-RAC}}
	\caption{(a) A maximally-dense $1$-planar graph with $5$ vertices. (b) A $1$-planar graph with $12$ vertices that is maximal but not optimal. (c) An optimal $1$-planar graph. (d) An optimal straight-line RAC graph.}\label{fi:density-definitions}
\end{figure}

The problem of establishing the maximum number of edges in a given type of beyond-planar graph has been extensively studied in the literature. We recall some basic definitions needed to describe the main results in this research direction. Let $\mathcal{F}$ be a beyond-planar graph family and let $G$ be an $n$-vertex graph in $\mathcal{F}$. $G$ is \emph{maximal} (in $\mathcal{F}$) if adding any edge to $G$ leads to a graph that is not in $\mathcal{F}$. $G$ is \emph{maximally dense} if it has the maximum number of edges over all $n$-vertex graphs in $\mathcal{F}$. The \emph{edge density} of $G$ is the ratio between its number of edges and its number of vertices. $G$ is \emph{optimal} if it has the maximum edge density over all graphs of $\mathcal{F}$. Note that an optimal graph is also maximally dense, while the converse may not be true; similarly, a maximally dense graph is maximal, but there are maximal graphs that are not maximally dense; Fig.~\ref{fi:maximally-dense} shows a maximally-dense $1$-planar graph with $5$ vertices that is not optimal; Fig.~\ref{fi:maximal} shows a $1$-planar graph with $12$ vertices that is maximal but not optimal, and Fig.~\ref{fi:optimal} shows an optimal $1$-planar graph with $12$ vertices.

\myparagraph{Density of $k$-planar graphs.} For $1 \le k \le 4$, Pach and T\'oth prove that $k$-planar graphs have at most $(k+3)(n-2)$ edges~\cite{DBLP:journals/combinatorica/PachT97}, and also use this result to improve an earlier lower bound on the crossing number of a graph. In particular, the resulting bounds are tight for $k=1$ and for $k=2$, which means that the optimal $1$-planar graphs have $4n-8$ edges and the optimal $2$-planar graphs have $5n-10$ edges. The best known upper bounds for $k=3$ and $k=4$ are $5.5n-11$~\cite{Pach2006} and $6n-12$~\cite{DBLP:journals/corr/Ackerman15}, respectively. For $k > 4$, the best known upper bound is $4.108\sqrt{k}n$~\cite{DBLP:journals/combinatorica/PachT97}. Lower bounds on the edge density of maximal $1$-planar graphs are given in~\cite{DBLP:conf/gd/BrandenburgEGGHR12}.  For bipartite $1$-planar graphs, Karpov proves a tight bound of $3n-6$ when $n$ is even and $n \neq 6$, and of $3n -9$ otherwise~\cite{Karpov2014}. Different subfamilies of $1$-planar graphs have also been studied, such as $1$-planar graphs where no two pairs of crossing edges share an end-vertex, called \emph{IC-planar graphs}, and those for which two pairs of crossing edges share at most one end-vertex, called \emph{NIC-planar graphs}. We report these bounds in Table~\ref{ta:density}. More results about the edge density of $1$-planar graphs are surveyed in~\cite{DBLP:journals/csr/KobourovLM17}.  

\myparagraph{Density of $k$-quasi planar graphs.} In 1996, Pach, Shahrokhi, and Szegedy conjectured that, for any fixed $k \ge 2$, there exists a constant $c_k$, depending only on $k$, such that every $k$-quasi planar graph on $n$ vertices has at most $c_k n$ edges. This conjecture has been proved for $k=3$ and $k=4$ by various authors, with the best upper bounds currently known being  $6.5n-20$~\cite{DBLP:journals/jct/AckermanT07} and $72(n-2)$~\cite{DBLP:journals/dcg/Ackerman09}, respectively. For $k>4$ the conjecture remains unproved, but several superlinear upper bounds have been established; the current best upper bound is $c_k n \log n$, for a suitable constant $c_k$~\cite{DBLP:journals/comgeo/SukW15}. An $O(n \log n)$ upper bound for $k$-quasi planar graphs with $x$-monotone edges is also known~\cite{DBLP:conf/gd/Valtr97,DBLP:journals/dcg/Valtr98}.

\myparagraph{Density of fan-planar, $k$-fan-crossing-free, and $k$-gap graphs.} Both fan-planar and $1$-gap planar graphs on $n$ vertices have at most $5n-10$ edges, which is a tight bound~\cite{DBLP:journals/corr/KaufmannU14,DBLP:conf/gd/Bae17}. Recall that the same bound holds for $2$-planar graphs. For $k > 1$, a $k$-gap planar graph has $O(\sqrt{k}n)$ edges~\cite{DBLP:conf/gd/Bae17}.
A fan-crossing-free graph has at most $4n-8$ edges, and at most $4n-9$ if the edges are drawn as straight lines~\cite{DBLP:journals/algorithmica/CheongHKK15}. These bounds are analogous to those for topological and geometric $1$-planar graphs. For $k>1$, $k$-fan-crossing-free graphs have at most $3(k-1)(n-2)$ edges~\cite{DBLP:journals/algorithmica/CheongHKK15}.

\myparagraph{Density of straight-line RAC graphs, ACE$\alpha$ graphs, and ACL$\alpha$ graphs.}
Since straight-line RAC graphs are fan-crossing-free, they cannot have more than $4n-9$ edges. In fact, the maximum number of edges of a straight-line RAC graph is $4n-10$~\cite{DBLP:journals/tcs/DidimoEL11}, and for each $k \geq 3$ there exists a straight-line RAC graph $G_k$ with $n=3k-5$ vertices and $4n-10$ edges (i.e., optimal). Graph $G_k$ is constructed as follows (see Fig.~\ref{fi:optimal-RAC}): $(i)$ start from a triangulated plane graph on $k$ vertices; $(ii)$ add to this graph its dual, except for the node corresponding to the external face; $(iii)$ for each node $u$ of the dual, add the three edges that connect $u$ to the three vertices of the face corresponding to $u$. The fact that two edges of a straight-line RAC drawing $\Gamma$ of a graph $G$ can cross only at right angles immediately implies that the \emph{crossing graph} of $\Gamma$ is bipartite; the crossing graph of $\Gamma$ has a vertex $v_e$ for each edge $e$ of $G$ and an edge $(v_e,v_{e'})$ if the $e$ and $e'$ cross in $\Gamma$. Hence, we can bi-color the edges of $G$ such that each color set induces a planar graph, which implies that $G$ has at most $6n-12$ edges. The $4n-10$ bound is proved by exploiting a finer coloring of the edges of $G$ with three colors and different counting arguments based on Euler's formula for planar graphs~\cite{DBLP:journals/tcs/DidimoEL11}. Dujmovic et al. provide an alternative proof of this bound, based on charging techniques~\cite{DBLP:journals/cjtcs/DujmovicGMW11}. They also show that straight-line ACL$\alpha$ graphs have at most $\frac{\pi}{\alpha}(3n-6)$ edges; the proof is based on a partition of the edges into distinct buckets, according to their directions, so that each bucket induces a planar graph. With similar arguments, it is shown that straight-line ACE$\alpha$ graphs have at most $3(3n-6)$ edges~\cite{aft-sgapd-10}. These bounds for ACL$\alpha$ and ACE$\alpha$ graphs are not known to be tight.

\myparagraph{Density of other graph families.} Since a skewness-$k$ graph $G$ becomes planar after the removal of at most $k$ edges, $G$ has at most $3n-6+k$ edges. This bound is trivially achieved by adding $k$ edges to a maximal planar graph. Similarly, removing at most $k$ vertices from an $n$-vertex $k$-apex graph yields a planar graph, and thus $k$-apex graphs have at most $3(n-k)-6+\sum_{i=1}^k(n-k+i-1)$ edges. A graph attaining this bound is constructed from a maximal planar graph, iteratively adding one vertex connected to all other vertices of the graph. $(k,l)$-grid-free and radial-$(k,l)$-grid-free graphs have at most $16 \cdot 24^{4^l}kn$ edges and $8 \cdot 24^lkn$ edges, respectively~\cite{DBLP:journals/gc/PachPST05}, while a bound $O(n \log^{4k-6} n)$ is proved for natural-$k$-grid-free graphs~\cite{DBLP:journals/comgeo/AckermanFPS14}. Concerning straight-line (3-length-path)-sif and straight-line (4-length-path)-sif graphs, it is shown that they have $O(n \log n)$ and $O(n\log n/\log\log n)$ edges, respectively~\cite{DBLP:journals/ejc/PachPTT04}, while straight-line (4-length-cycle)-sif graphs have $O(n^{8/5})$ edges~\cite{DBLP:conf/compgeom/PinchasiR03}. Finally, planarly-connected graphs have $c\cdot n$ edges, where $c$ is an absolute constant~\cite{DBLP:journals/jgaa/Ackerman17}.

\smallskip Table~\ref{ta:density} summarizes the discussed bounds on the edge density of beyond-planar graphs. 
%Additional bounds for constrained scenarios are reported in Section~\ref{se:constraints}.

\begin{table}
\centering
\caption{Density. Most of the bounds in the table hold for simple drawings only; while few of them are still valid for non-simple drawings. The symbol $\times$ in column \textsc{Tightness} means that either the bound is far from being tight or that it is tight only up to a multiplicative or additive constant.\label{ta:density}}\smallskip{%
\renewcommand{\arraystretch}{1.3}
\footnotesize
%\resizebox{\textwidth}{!}{
\begin{tabular}{|l|c|c|l|}

\hline
 \textsc{Graph Family} & \textsc{Max. Num. Edges} & \textsc{Tight} & \textsc{References} \\\hline
 {$1$-planar} & {$4n - 8$} & {$\checkmark$} & {\cite{BSW83,DBLP:journals/combinatorica/PachT97}}  \\\hline
 {straight-line $1$-planar} & {$4n - 9$} & {$\checkmark$} & {\cite{D13,Ackerman14}}  \\\hline
 {bipartite $1$-planar} & \makecell{\parbox[c][0.8cm]{2.6cm}{$3n-8$ for even $n \neq 6$ \\ $3n-9$ otherwise}} & {$\checkmark$} & {\cite{Karpov2014}}  \\\hline
 {IC-planar} & {$3.25n - 6$} & {$\checkmark$} & {\cite{Zhang2013}}  \\\hline
 {NIC-planar} & {$3.6n - 7.2$} & {$\checkmark$} & {\cite{CzapS17}}  \\\hline
 {$2$-planar} & {$5n - 10$} & {$\checkmark$} & {\cite{DBLP:journals/combinatorica/PachT97}}  \\\hline
 %{$3$-planar} & {$5.5(n - 2)$} & {$\times$\footnote{Tight nel caso non semplice. Altrimenti $5.5n-15$.}} & {\cite{Pach2006}}  \\\hline
 {$3$-planar} & {$5.5(n - 2)$} & {$\times$} & {\cite{Pach2006}}  \\\hline
 %{$4$-planar} & {$6n - 12$} & {$\times$\footnote{$6n-18$.}} & {\cite{DBLP:journals/corr/Ackerman15}}  \\\hline
 {$4$-planar} & {$6n - 12$} & {$\times$} & {\cite{DBLP:journals/corr/Ackerman15}}  \\\hline
 {$k$-planar $(k \geq 5)$} & {$4.108\sqrt{k}n$} & {$\times$} & {\cite{DBLP:journals/combinatorica/PachT97}}  \\\hline
 {$3$-quasi planar} & {$6.5n-20$} & {$\times$} & {\cite{DBLP:journals/jct/AckermanT07}}  \\\hline
 {$4$-quasi planar} & {$72(n-2)$} & {$\times$} & {\cite{DBLP:journals/dcg/Ackerman09}}  \\\hline
 {$k$-quasi planar $(k \geq 5)$} & {$c_k n \log n$} & {$\times$} & {\cite{DBLP:journals/comgeo/SukW15}}  \\\hline
 {skewness-$k$} & {$3n-6+k$} & {$\checkmark$} & {Trivial}  \\\hline
 {$k$-apex} & {$3(n-k)-6+\sum_{i=1}^k(n-k+i-1)$} & {$\checkmark$} & {Trivial}  \\\hline
 {$(k,l)$-grid-free} & {$16 \cdot 24^{4^l}kn$} & {$\times$} & {\cite{DBLP:journals/gc/PachPST05}} \\\hline
 {radial-$(k,l)$-grid-free} & {$8 \cdot 24^lkn$} & {$\times$} & {\cite{DBLP:journals/gc/PachPST05}} \\\hline
 {natural-$k$-grid-free} & {$O(n \log^{4k-6} n)$} & {$\times$} & {\cite{DBLP:journals/comgeo/AckermanFPS14}} \\\hline
 {straight-line natural-$k$-grid-free} & {$O(kn \log^2 n)$} & {$\times$} & {\cite{DBLP:journals/comgeo/AckermanFPS14}} \\\hline
 {fan-crossing-free} & {$4n-8$} & {$\checkmark$} & {\cite{DBLP:journals/algorithmica/CheongHKK15}}  \\\hline
 {straight-line fan-crossing-free} & {$4n-9$} & {$\checkmark$} & {\cite{DBLP:journals/algorithmica/CheongHKK15}}  \\\hline
 {$k$-fan-crossing-free ($k \ge 3$)} & {$3(k-1)(n-2)$} & {$\times$} & {\cite{DBLP:journals/algorithmica/CheongHKK15}}  \\\hline
 {fan-planar} & {$5n-10$} & {$\checkmark$} & {\cite{DBLP:journals/corr/KaufmannU14}}  \\\hline
 {straight-line RAC} & {$4n - 10$} & {$\checkmark$} & {\cite{DBLP:journals/tcs/DidimoEL11}}  \\\hline
 {straight-line ACE$\alpha$} & {$3(3n-6)$} & {$\times$} & {\cite{aft-sgapd-10}}  \\\hline
 {straight-line ACL$\alpha$} & {$\frac{\pi}{\alpha}(3n-6)$} & {$\times$} & {\cite{DBLP:journals/cjtcs/DujmovicGMW11}}  \\\hline
 {$1$-gap-planar} & {$5n-10$} & {$\checkmark$} & {\cite{DBLP:conf/gd/Bae17}}  \\\hline
 {$k$-gap-planar ($k>1$)} & {$O(\sqrt{k}n)$} & {$\times$} & {\cite{DBLP:conf/gd/Bae17}}  \\\hline
 {straight-line (3-length-path)-sif} & {$O(n \log n)$} & {$\times$} & {\cite{DBLP:journals/ejc/PachPTT04}} \\\hline
 {straight-line (5-length-path)-sif} & {$O(n\log n/\log\log n)$} & {$\times$} & {\cite{DBLP:journals/ejc/PachPTT04}} \\\hline
 {straight-line (4-length-cycle)-sif} & {$O(n^{8/5}$)} & {$\times$} & {\cite{DBLP:conf/compgeom/PinchasiR03}} \\\hline
 {planarly-connected} & {$c n$} & {$\times$} & {\cite{DBLP:journals/jgaa/Ackerman17}} \\\hline
\end{tabular}
%}
}
%\begin{tabnote}%
%\Note{Notes.}{Most of the bounds in the table hold for simple drawings only; while few of them are still valid for non-simple drawings. The symbol $\times$ in column \textsc{Tightness} means that either the bound is far from being tight or that it is tight only up to a multiplicative or additive constant.}
%\end{tabnote}
\end{table}% 

%\subsection{Open Problems}\label{ss:ed-open}

\myparagraph{Open Problems.} Although Tur\'an-type questions on beyond-planar drawings have been studied for a long time and the geometric graph theory literature is reach of results, there are still many beautiful open problems. For several families of beyond-planar graphs the upper bounds on the maximum number of edges reported in Table~\ref{ta:density} are not tight. Hence, the goal of achieving tight bounds for each of these families gives rise to an array of challenging problems. In particular, we find of particular interest the following (see also \cite{DBLP:books/daglib/0017422,DBLP:reference/cg/2017}).

\begin{problem}\label{op:pss-conjecture} Find a tight upper bound on the edge density of $k$-quasi planar graphs for $k>4$. The problem is relevant also when the edges are drawn as $x$-monotone curves.
\end{problem}

\begin{problem}\label{op:tight-density-bounds-k-planar} Find tight upper bounds on the edge density of (simple) $k$-planar graphs for $k \geq 3$.
\end{problem}

Starting references for Problem~\ref{op:pss-conjecture} include~\cite{DBLP:journals/siamdm/FoxPS13,DBLP:journals/comgeo/SukW15,DBLP:conf/gd/Valtr97,DBLP:journals/dcg/Valtr98}. We recall that Pach et al. conjecture an upper bound that is linear in the number of vertices~\cite{DBLP:journals/algorithmica/PachSS96}. 
Concerning Problem~\ref{op:tight-density-bounds-k-planar}, recent papers that discuss the edge density of (not necessarily simple) $3$-planar graphs include~\cite{DBLP:conf/gd/Bekos0R16,DBLP:conf/compgeom/Bekos0R17}.

Another research direction is concerned with providing lower bounds on the number of edges of maximal beyond-planar graphs. This type of question is mostly unexplored; results are known for $1$- and $2$-planar graphs~\cite{DBLP:conf/gd/AuerBGH12,DBLP:conf/gd/BrandenburgEGGHR12}. We suggest the following.

\begin{problem}\label{op:density-lower-bound-maximal-RAC} Find a lower bound on the edge density of maximal straight-line RAC graphs.
\end{problem}

%\begin{problem}\label{op:tight-density-bounds-RAC} Provide tight upper bounds on the maximum number of edges for the following families of beyond-planar drawings: $k$-bend RAC drawings, for $k \in \{1,2\}$; straight-line ACL$\alpha$ drawings; $k$-bend ACE$\alpha$ drawings, for $k\in\{1,2,3\}$.
%\end{problem}

%See also~\cite{dl-13} for references about Problem~\ref{op:tight-density-bounds-RAC}.

%
\section{Recognition}\label{se:recognition}
Given a graph $G$ and a family $\mathcal F$ of beyond-planar graphs, the recognition problem studies the complexity of deciding  whether a graph $G$ belongs to $\mathcal F$. As we have seen in Section~\ref{se:density}, several families of beyond-planar graphs are sparse and this may suggest a correspondence with planar graphs, which can be recognized in linear-time~\cite{DBLP:journals/jacm/HopcroftT74}. Unfortunately, this is not the case, and for most of the studied beyond-planar graph families the recognition problem is in fact hard.

Recognizing $1$-planar graphs is \npc in general~\cite{GB07,KM13} and it remains \npc even if the skeweness of the input graph is $1$~\cite{DBLP:journals/algorithmica/CabelloM11}. The problem is however fixed-parameter tractable with respect to the vertex-cover number, the cyclomatic number, or the tree-depth of the input graph~\cite{JGAA-457}. Recognizing $1$-planar graphs remains \npc also when the input graph comes with a fixed rotation system, which must be preserved~\cite{AuerBGR15}. On the positive side, deciding whether a graph with $n$ is optimal $1$-planar (i.e., it is $1$-planar and has $4n-8$ edges) is $O(n)$-time solvable~\cite{Brandenburg16a}; the testing algorithm exploits a structural characterization of optimal $1$-planar graphs~\cite{DBLP:journals/dm/Suzuki10}. We refer the reader to the annotated bibliography by Kobourov et al. for further references on recognizing meaningful subclasses of $1$-planar graphs~\cite{DBLP:journals/csr/KobourovLM17}. It is worth remarking that several papers present interesting bounds of graph parameters for $1$-planar and $k$-planar graphs, which shed some light on the structure of these graphs and thus which may be of interest for the design of fixed-parameter tractable recognition algorithms. For example, $k$-planar graphs on $n$ vertices have $O(\log n)$ book thickness~\cite{JGAA-454}, $O(\sqrt{kn})$ treewidth and $O(k)$ layered treewidth~\cite{DBLP:journals/siamdm/DujmovicEW17}, and bounded expansion~\cite{NesetrilMW12} (see ~\cite{DBLP:journals/csr/KobourovLM17} for more results).

Recognizing skewness-$k$ graphs is \npc in the general case~\cite{liu79}, but the problem can be solved in $O(n)$ time for any fixed value of $k$ (i.e., it is FPT with parameter $k$)~\cite{Kawarabayashi:2007:CCN:1250790.1250848}. 

The set of $1$-apex graphs is closed under the operation of taking minors~\cite{DBLP:conf/focs/GuptaI91}, hence these graphs have a forbidden graph characterization by the Robertson-Seymour theorem~\cite{DBLP:journals/jct/RobertsonS04}. However, the set of obstructions for these graphs has only been partially discovered~\cite{pierce2014}. Recognizing $n$-vertex $k$-apex graphs is \npc~\cite{DBLP:journals/jcss/LewisY80}, but there is an $O(n)$-time algorithm if $k$ is a fixed constant~\cite{DBLP:conf/focs/Kawarabayashi09}.

Recognizing fan-planar graphs is \npc~\cite{DBLP:journals/tcs/BinucciGDMPST15}, even for fixed rotation system~\cite{bcghk-16}. The same  holds for $1$-gap-planar graphs~\cite{DBLP:conf/gd/Bae17}.

Recognizing straight-line RAC drawable graphs is \nph in general~\cite{DBLP:journals/jgaa/ArgyriouBS12}, while it can be solved in linear time for complete bipartite graphs~\cite{DBLP:journals/ipl/DidimoEL10}. As recently shown, even the more restricted problem of deciding whether a graph admits a straight-line RAC drawing with at most one crossing per edge is \nph~\cite{Bekos2017}. It is however unknown if any of these two problems belongs to \np, because it is unknown the complexity of deciding whether an embedded graph admits an embedding-preserving straight-line RAC ($1$-planar) drawing. On the other hand, every $1$-plane graph admits a $1$-bend RAC drawing with at most one crossing per edge~\cite{Bekos2017,DBLP:conf/ewcg/Chaplick18} (see also Section~\ref{se:aesthetics}).

Concerning $\alpha$-SHPEDs, there are necessary or sufficient conditions for their existence, although the complexity of the recognition problem has not been established. If an $n$-vertex complete graph has a $\frac{1}{4}$-SHPED, then $n<165$; also, if a graph has bandwidth $k$, it has a $\Theta(\frac{1}{k})$-SHPED~\cite{JGAA-438}. Similar bounds are given for complete bipartite graphs. In~\cite{JGAA-438} it is also studied the problem of maximizing the total stub length (or ink), so to turn a geometric graph into a partial edge drawing that is symmetric but not necessarily homogeneous (the value of $\alpha$ may not be the same for all the edges). This problem is \nph, but becomes polynomial-time solvable for geometric $2$-planar graphs.

\smallskip Table~\ref{ta:recognition} summarizes the results discussed in this section.

\begin{table}
\centering
\caption{Recognition.\label{ta:recognition}}\smallskip{%
\renewcommand{\arraystretch}{1.3}
\footnotesize
%\resizebox{\textwidth}{!}{
\begin{tabular}{|l|l|c|l|}
\hline
 \textsc{Graph Family} & \textsc{Restrictions} & \textsc{Complexity} & \textsc{References} \\\hline
 {$1$-planar} & {} & {\npc} & {\cite{GB07,KM13}}\\\hline
 {$1$-planar} & {bounded bandwidth, pathwidth, or treewidth} & {\npc} & {\cite{JGAA-457}}\\\hline
 {$1$-planar} & {bounded cyclomatic number or treedepth} & {$O(n)$} & {\cite{JGAA-457}}\\\hline
 {$1$-planar} & {fixed rotation system} & {\npc} & {\cite{AuerBGR15}}\\\hline
 {$1$-planar} & {skewness-$1$} & {\npc} & {\cite{DBLP:journals/siamcomp/CabelloM13}}\\\hline
 {optimal $1$-planar} & {} & {$O(n)$} & {\cite{Brandenburg16a}}\\\hline
 {skewness-$k$} & {} & {\npc} & {\cite{liu79}}\\\hline
 {skewness-$k$} & {bounded $k$} & {$O(n)$} & {\cite{Kawarabayashi:2007:CCN:1250790.1250848}}\\\hline
 {$k$-apex} & {} & {\npc} & {\cite{DBLP:journals/jcss/LewisY80}}\\\hline
 {$k$-apex} & {bounded $k$} & {$O(n)$} & {\cite{DBLP:conf/focs/Kawarabayashi09}}\\\hline
 {fan-planar} & {} & {\npc} & {\cite{DBLP:journals/tcs/BinucciGDMPST15}}\\\hline
 {fan-planar} & {fixed rotation system} & {\npc} & {\cite{bcghk-16}}\\\hline
 {straight-line RAC} & {} & {\nph} & {\cite{DBLP:journals/jgaa/ArgyriouBS12}}\\\hline
 {staight-line $1$-planar RAC} & {} & {\nph} & {\cite{Bekos2017}}\\\hline
 {$1$-gap-planar} & {} & {\npc} & {\cite{DBLP:conf/gd/Bae17}}\\\hline
 {$1$-gap-planar} & {fixed rotation system} & {\npc} & {\cite{DBLP:conf/gd/Bae17}}\\\hline
\end{tabular}
%}
}
\end{table}%

%\subsection{Open Problems}

\myparagraph{Open Problems.} As Table~\ref{ta:recognition} shows, the recognition problem remains unexplored for many families of beyond-planar graphs. This poses several open questions. For example:

\begin{problem}\label{op:k-qp}
What is the complexity of deciding whether a graph is $k$-quasi planar?  The question is already interesting when $k=3$.
\end{problem}

It is know that every graph has a RAC drawing with at most three bends per edge~\cite{DBLP:journals/tcs/DidimoEL11} and that it has $O(n)$ edges if either one bend or two bends per edge are allowed~\cite{DBLP:journals/comgeo/ArikushiFKMT12} (see Section~\ref{se:aesthetics}). Recognizing straight-line RAC graphs is \nph~\cite{DBLP:journals/jgaa/ArgyriouBS12}, but the following question is still open.

\begin{problem}\label{op:1-2bRAC}
What is the complexity of recognizing whether a graph admits a $k$-bend RAC drawing, for $k \in\{1,2\}$?
\end{problem}

A characterization of the graphs that admit an $\alpha$-SHPED is unknown.

\begin{problem}\label{op:SHPED}
What is the complexity of deciding whether a graph admits an $\alpha$-SHPED? In particular, what is the complexity for $\alpha = \frac{1}{4}$? 
\end{problem}

\section{Stretchability}\label{se:stretchability}
\begin{figure}[t]
	\centering
	\subfigure[]{\includegraphics[scale=0.8,page=1]{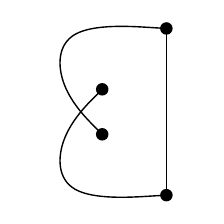}\label{fi:b-conf}}\hfill
	\subfigure[]{\includegraphics[scale=0.8,page=2]{figs/badconf}\label{fi:w-conf}}\hfill
	\subfigure[]{\includegraphics[scale=0.6,page=2]{figs/sl-drawing}\label{fi:sl-drawing-2}}\hfill
	\subfigure[]{\includegraphics[scale=0.6,page=3]{figs/sl-drawing}\label{fi:sl-drawing-3}}
	\caption{(a) B-configuration. (b) W-configuration. (c) A $1$-plane graph $G$. (d) An embedding-preserving straight-line drawing of $G$.}
\end{figure}

The famous F{\'a}ry's theorem states that every plane graph is \emph{stretchable}, i.e., it has an embedding-preserving straight-line drawing~\cite{fary-48} (this result was also independently proven by Wagner~\cite{wagner-36} and Stein~\cite{stein-51}). So far, limited effort has been done to extend this result in the context of graph drawing beyond planarity.
In 1988, Thomassen initiated the study of the stretchability problem for $1$-plane graphs~\cite{th-rdg-88}. He proved that, differently from plane graphs, a $1$-plane graph admits an embedding-preserving straight-line drawing if and only if it contains neither \emph{B-configurations} nor \emph{W-configurations} as subgraphs~\cite{th-rdg-88} (see Figs.~\ref{fi:b-conf}-\ref{fi:w-conf}).
Later on, Thomassen's characterization has been used to design a linear-time algorithm to test whether a $1$-plane graph is stretchable~\cite{DBLP:conf/cocoon/HongELP12}. This algorithm is based on an efficient procedure that checks whether a $1$-plane graph $G$ contains any B- or W-configuration, and, if not, it also returns a straight-line drawing of $G$. Figs.~\ref{fi:sl-drawing-2}-\ref{fi:sl-drawing-3} show a stretchable $1$-plane graph $G$ and a straight-line drawing of $G$, respectively. Recently, Hong and Nagamochi have studied the stretchability problem of $1$-plane graphs in a more relaxed setting~\cite{HongN16}. They describe a linear-time algorithm that tests whether a $1$-plane graph is stretchable assuming that the rotation system and the external face of the graph can change, while it is required that the pairs of crossing edges stay the same.
Eades et al. study the stretchabilty problem for skewness-$1$ graphs~\cite{DBLP:conf/wads/EadesHLKP15}. They characterize the maximal topological skewness-$1$ graphs that are stretchable, and give a linear-time testing and drawing algorithm based on this characterization.

Table~\ref{ta:stretchability} summarizes the aforementioned results.

\begin{table}[]%
\centering
	\caption{Stretchability.\label{ta:stretchability}}\smallskip{%
		\renewcommand{\arraystretch}{1.3}
		\footnotesize
		%\resizebox{\textwidth}{!}{
			\begin{tabular}{|l|l|c|l|}
				\hline
				\textsc{Graph Family} & \textsc{Restrictions} & \textsc{Complexity of testing} & \textsc{References} \\\hline
				{$1$-planar} & {fixed embedding} & {$O(n)$} & {\cite{DBLP:conf/cocoon/HongELP12}}\\\hline
				{$1$-planar} & {fixed pairs of crossing edges} & {$O(n)$} & {\cite{HongN16}}\\\hline
				{maximal skewness-$1$} & {fixed embedding} & {$O(n)$} & {\cite{DBLP:conf/wads/EadesHLKP15}}\\\hline
			\end{tabular}
		%}
	}
\end{table}%

%\subsection{Open Problems}

\myparagraph{Open Problems.} The stretchability problem for beyond-planar graphs is a fertile and essentially unexplored research subject. As a matter of fact, for any type of forbidden crossing configuration, the corresponding stretchability question gives rise to an open problem. For example:

\begin{problem}\label{op:str-qp}
Characterize the $k$-quasi planar topological graphs that are stretchable. This question is interesting also when $k=3$.
\end{problem}

Another research stream is about embedding-preserving drawings with good crossing angle resolution. This question is interesting even for structurally simple topological graphs. For example:

\begin{problem}\label{op:str-rac}
Does every topological graph of maximum vertex degree three admit an embedding-preserving straight-line RAC drawing?
\end{problem}

Finally, we recall that a characterization of the almost-plane graphs that are stretchable is known only when the number of edges is $3n-5$ (a characterization for fewer edges is known on the sphere but not in the plane)~\cite{DBLP:conf/wads/EadesHLKP15}. Hence, a natural question is the following.

\begin{problem}\label{op:str-ap}
Characterize the topological skewness-$k$ graphs that are stretchable. This question is interesting also when $k=1$ and the number of edges is smaller than $3n-5$.
\end{problem}

\section{Relationships Between Graph Families}\label{se:relationships}
Some families of beyond-planar graphs have similar edge densities or exhibit similar structural and topological properties. In these cases it is natural to ask whether they have some inclusion relationships. In what follows, we survey the main results concerning this research direction.

\begin{figure}[t]
	\centering
	\subfigure[]{\includegraphics[width=0.23\columnwidth,page=1]{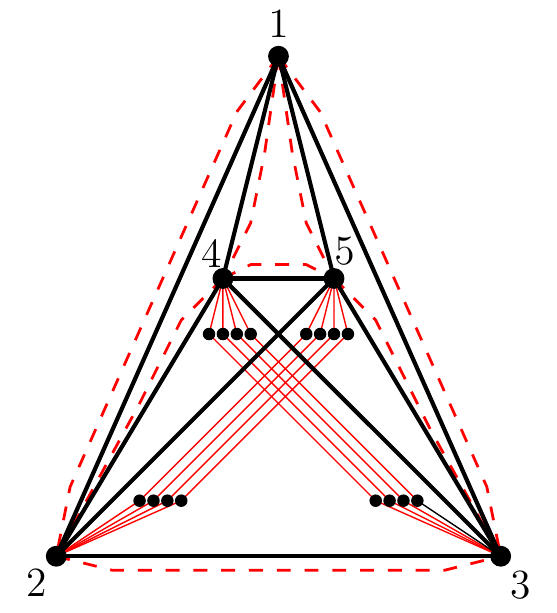}\label{fi:RAC-not-1-planar}}\hfill
	\subfigure[]{\includegraphics[width=0.23\columnwidth,page=2]{figs/relationship}\label{fi:1-planar-not-RAC}}\hfill
	\subfigure[]{\includegraphics[width=0.23\columnwidth,page=3]{figs/relationship}\label{fi:1-planar-1-bend-RAC}}	
	\caption{(a) Schematic illustration of a straight-line RAC graph that is not $1$-planar; each dashed edge $(u,v)$ represents four paths of three edges between $u$ and $v$, like those connecting vertices $\{3, 4\}$ and vertices $\{2, 5\}$. (b) A $1$-plane graph with $4n-10$ edges that does not admit a straight-line RAC drawing; an infinite family of this type of graphs can be obtained by recursively plugging the whole graph inside the 4-cycle $\{5,6,7,8\}$. (c) A $1$-planar $1$-bend RAC drawing of the graph in Figure~\ref{fi:1-planar-not-RAC}.}\label{fi:relationship}
\end{figure}

\myparagraph{RAC graphs and $1$-planar graphs.} One of the most studied problems on the subject is the relationship between RAC graphs and $1$-planar graphs. Recall that straight-line RAC drawings have at most $4n - 10$ edges, while topological (resp. geometric) $1$-planar graphs have at most $4n-8$ edges (resp. $4n-9$). This immediately implies that optimal $1$-planar graphs are not straight-line RAC. Eades and Liotta proved in fact that these two families are in general incomparable, and provide a series of interesting results about their relationships~\cite{DBLP:journals/dam/EadesL13}. They exhibit an infinite subfamily of straight-line RAC graphs with $n \geq 85$ vertices that are not $1$-planar (see, e.g., Fig.~\ref{fi:RAC-not-1-planar}), and they show that there exist infinitely many $1$-planar graphs with $4n-10$ edges that are not straight-line RAC (see, e.g., Fig.~\ref{fi:1-planar-not-RAC}). On the positive side, every optimal straight-line RAC graph is $1$-planar (see, e.g., Fig.~\ref{fi:optimal-RAC}).

Brandenburg et al. study the RAC drawability of the $1$-planar graphs with independent pairs of crossing edges (the IC-planar graphs)~\cite{DBLP:journals/tcs/BrandenburgDEKL16}. They show that every IC-planar graph has a straight-line RAC drawing, while this is not true for the larger class of the NIC-planar graphs~\cite{DBLP:journals/dam/BachmaierBHNR17}.

The incomparability of straight-line RAC graphs and $1$-planar graphs, together with the fact that IC-planar graphs always admit a straight-line RAC drawing (which is also $1$-planar), suggest two interesting questions: $(i)$ What is the complexity of deciding whether a graph admits a drawing that is both $1$-planar and straight-line RAC? $(ii)$ Does every $1$-planar graph admit a $1$-planar drawing that is also RAC if we allow at most one bend per edge? These questions have been recently answered~\cite{Bekos2017}: Question $(i)$ is \nph, as already mentioned in Section~\ref{se:recognition} (see Table~\ref{ta:recognition}), while Question $(ii)$ has a positive answer. Figure~\ref{fi:1-planar-1-bend-RAC} shows a $1$-planar RAC drawing with at most one bend per edge of the graph in Fig.~\ref{fi:1-planar-not-RAC}. 
Additional inclusion relationships between $1$-planar and straight-line RAC drawings have been studied for constrained drawings and they are summarized in Section~\ref{se:constraints}.

\myparagraph{$k$-planar graphs and $k$-quasi planar graphs.}  Note that the tight bounds on the edge density of $3$-planar and $3$-quasi planar graphs imply the existence of $3$-quasi planar graphs that are not $3$-planar. On the other hand, for $k \ge 1$, every $k$-planar graph is clearly $(k + 2)$-quasi planar. These two observations have motivated a recent work on the relationship between $k$-planar graphs and $k$-quasi planar graphs~\cite{DBLP:conf/wg/AngeliniBBLBDLM17}. This work proves that, for $k \ge 3$, every $k$-planar graph is $(k+1)$-quasi planar. The proof is based on a rerouting argument that starts from a $k$-planar drawing and resolves all possible bundles of $k+1$ pairwise crossing edges. The drawing produced by this technique is $(k+1)$-quasi planar, but it may not be $k$-planar anymore. This result has been later extended to the case $k=2$ ~\cite{DBLP:conf/mfcs/0001T17}. Thus, every $k$-planar graph is $(k+1)$-quasi planar, for $k \ge 2$.

\myparagraph{$k$-planar graphs and fan-planar graphs.} Since optimal fan-planar graphs have the same density as optimal $2$-planar graphs (optimal $n$-vertex graphs have $5n-10$ edges for these two families),  Binucci et al. study the relationship between fan-planar and $k$-planar graphs~\cite{DBLP:journals/tcs/BinucciGDMPST15}. They prove that these two families are incomparable. On the one hand, they show that for any $k \geq 2$ there exists a fan-planar graph that is not $k$-planar; the proof uses a complete $3$-partite graph $K_{1,3,h}$, where the index $h$ depends on $k$; this graph is clearly fan-planar but any of its drawings contains too many crossings to be $k$-planar. On the other hand, they exhibit $2$-planar graphs that are not fan-planar.

\myparagraph{$k$-gap planar graphs, $k$-planar graphs, $k$-quasi planar graphs.} Bae et al. studied the relationship between $k$-gap planar graphs and both $k$-planar and $k$-quasi planar graphs~\cite{DBLP:conf/gd/Bae17}. By using Hall's theorem, they prove that for every $k \ge 1$ all $2k$-planar graphs are $k$-gap planar. On the other hand, for every fixed $k \ge 1$, there exists a $1$-gap planar graph that is not $k$-planar. Similarly, by using a counting argument on the number of crossings, they prove that all $k$-gap planar graphs are $2k+2$-quasi planar, while for all $k \ge 1$ they exhibit a quasi planar graph that is not $k$-gap planar.

\myparagraph{Planarly-connected graphs, $1$-planar graphs, fan-planar graphs.} Ackerman motivates the study of planarly-connected graphs with the fact that both maximally dense $1$-planar graphs and maximally dense fan-planar graphs are planarly-connected~\cite{DBLP:journals/jgaa/Ackerman17}. For $1$-planar graphs, the high-level idea is to consider a drawing with the minimum number of crossings of a maximally dense graph; in such a drawing, for every pair of crossing edges there is a crossing-free edge that connects their endpoints, as otherwise one could contradict either the fact that the drawing is crossing minimal or the fact that the graph is maximally dense. The idea for fan-planar graphs is similar.

\smallskip The intersection relationships between the different families of beyond-planar graph families are summarized in Table~\ref{ta:relationships}. Each family in a row (resp. column) is represented by a thin (resp. thick) circle. Each cell reports the relationship between the corresponding row and column using the common Venn diagram notation. Note that, some relationships immediately derive from the definitions. For example, a fan-planar drawing is always $k$-quasi planar (for each $k \geq 3$), because three mutually crossing edges imply that two independent edges are crossed by a third one.

\newcommand{\tfig}[1]{\setlength\fboxrule{0pt}\fbox{\includegraphics[page=#1]{figs/venn}}}

\begin{table}[h!]%
\caption{Relationships. In each pairwise comparison, when applicable, we assume $k \le h$. The table does not report inclusion relationships about restricted subfamilies, such as IC-planar graphs, NIC-planar graphs, and optimal straight-line RAC graphs. These relationships are discussed in the text.\label{ta:relationships}}\smallskip{%
\renewcommand{\arraystretch}{1.3}
\footnotesize
\resizebox{\textwidth}{!}{
\begin{tabular}{|c|c|c|c|c|c|c|c|c|}
\hline
 \textsc{
\diagbox{\tfig{1}}{\tfig{6}}} & \textsc{$1$-planar} & \textsc{$h$-planar $h \ge 2$} & \textsc{$h$-quasi pl. $h \ge 3$} & \textsc{fan planar} & \textsc{straight-line RAC} & \textsc{$h$-gap planar $h \ge 1$} & \textsc{planarly-conn.}\\\hline
 {\textsc{$1$-planar}} & {\tfig{7}} & {\tfig{5}} & {\tfig{5}} & {\tfig{5}} & {\tfig{2}} & {\tfig{5}} & {\tfig{2}} \\\hline
 {\textsc{$k$-planar $k \ge 2$}}& {\tfig{4}} & {\tfig{5}} & {\tfig{5} $h \ge k+1$} & {\tfig{2} $k=2$} & {\tfig{2}} & {\tfig{5} $k=2h$} & {\tfig{2}}  \\\hline
 {\textsc{$k$-quasi pl. $k \ge 3$}} & {\tfig{4}} & {\tfig{2}} & {\tfig{5}} & {\tfig{4}} & {\tfig{4}} & {\tfig{2} $k=3$} & {\tfig{2}}  \\\hline
 {\textsc{fan planar}} & {\tfig{4}} & {\tfig{2} $h=2$} & {\tfig{5}} & {\tfig{7}} & {\tfig{2}} & {\tfig{2}} & {\tfig{2}}  \\\hline
 {\textsc{straight-line RAC}} & {\tfig{2}} & {\tfig{2}} & {\tfig{5}} & {\tfig{2}} & {\tfig{7}} & {\tfig{2}} & {\tfig{2}}  \\\hline
 {\textsc{$k$-gap planar $k \ge 1$}} & {\tfig{4}} & {\tfig{2} $k=1$} & {\tfig{5} $h=2k+2$} & {\tfig{2}} & {\tfig{2}} & {\tfig{5}} & {\tfig{2}}  \\\hline
 {\textsc{planarly-conn.}} & {\tfig{2}} & {\tfig{2}} & {\tfig{2}} & {\tfig{2}} & {\tfig{2}} & {\tfig{2}} & {\tfig{7}}  \\\hline
\end{tabular}
}
}
%\begin{tabnote}%
%\Note{Notes.}{In each pairwise comparison, when applicable, we assume $k \le h$. The table does not report inclusion relationships about restricted subfamilies, such as IC-planar graphs, NIC-planar graphs, and optimal straight-line RAC graphs. These relationships are discussed in the text.}
%\end{tabnote}
\end{table}%

%\subsection{Open Problems}
\myparagraph{Open Problems.} Table~\ref{ta:relationships} shows that some pairs of beyond-planar graph families have a non-empty intersection that is not an inclusion. In these cases, it is interesting to characterize the graphs that belong to both families. For example, there are 1-planar graphs that are not straight-line RAC, there are straight-line RAC graphs that are not 1-planar, all optimal straight-line RAC graphs are 1-planar, and all IC-planar graphs are straight-line RAC~\cite{DBLP:journals/dam/EadesL13,DBLP:journals/tcs/BrandenburgDEKL16}. However, the following is still open.

\begin{problem}\label{op:RAC-1pl}
Characterize the straight-line RAC graphs that are 1-planar.
\end{problem}

Other questions can be asked for those cells of Table~\ref{ta:relationships} that show a proper inclusion between two families of beyond-planar graphs. For example, while it is known that every $k$-planar graph is $(k+1)$-quasi planar~\cite{DBLP:conf/wg/AngeliniBBLBDLM17,DBLP:conf/mfcs/0001T17} and that optimal $3$-planar graphs are known to be $3$-quasi planar~\cite{DBLP:conf/compgeom/Bekos0R17}, some fundamental questions on the combinatorial relationships between $k$-planarity and $h$-quasi planarity remain unanswered. For example:

\begin{problem}\label{op:k-pl-kquasi}
Is a $k$-planar graph also $k$-quasi planar? For sufficiently large values of $k$, is every $k$-planar graph $f(k)$-quasi planar, for some function $f(k)=o(k)$?
\end{problem}

%\begin{problem}\label{op:k-pl-kquasi.2}
% For sufficiently large values of $k$, is every $k$-planar graph $f(k)$-quasi planar, for some function $f(k)=o(k)$?
%\end{problem}

Finally, the relationships between some pairs of beyond-planar graph families have not yet been studied. For example:

\begin{problem}\label{op:gap-fan}
What is the relationship between $k$-gap planar and fan-planar graphs, for $k \geq 1$?
\end{problem}

\section{Aesthetics}\label{se:aesthetics}

While beyond-planar graphs focus on properties of edge crossings, other aesthetic requirements have been extensively used in the literature in order to produce geometric representations of graphs that are clear and pleasing for the reader. For example, in a polyline drawing, the vertices of the graph are points and the edges are polylines whose complexity should be kept as low as possible to assist the reader. A common measure to capture the edge complexity of a polyline drawing is the number of bends per edge. Results concerning the edge complexity of beyond-planar graphs are in Section~\ref{ss:Edge-Complexity}.  Other drawing paradigms, like visibility representations and contact representations, convey edges as horizontal or vertical segments and shift the complexity to the vertices, which are drawn using more general shapes such as bars, rectangles, or polygons. We survey work related to the vertex complexity of beyond-planar graphs in Section~\ref{ss:Vertex-Coplexity}. Finally,  a common goal for polyline drawings is to use integer coordinates for vertices and bend points and to fit the drawing into a small bounding box, so to display the drawing onto a screen with finite resolution. In particular, the \emph{area} of a polyline drawing is the area occupied by its \emph{bounding box}, i.e., by the minimum axis-aligned box containing the drawing. If the bounding box has side lengths $X-1$ and $Y-1$, we say that the drawing has area $X \times Y$. Results concerning the area requirement of beyond-planar graphs in combination with edge/vertex complexity requirements are considered in Sections~\ref{ss:Edge-Complexity}$-$\ref{ss:Vertex-Coplexity}. Furthermore, Section~\ref{ss:tradeoffs-planar} discusses results about area-crossings trade-offs for beyond-planar drawings of planar graphs.

\subsection{Edge Complexity}\label{ss:Edge-Complexity}

\begin{figure}
\centering
\includegraphics[width=0.8\columnwidth]{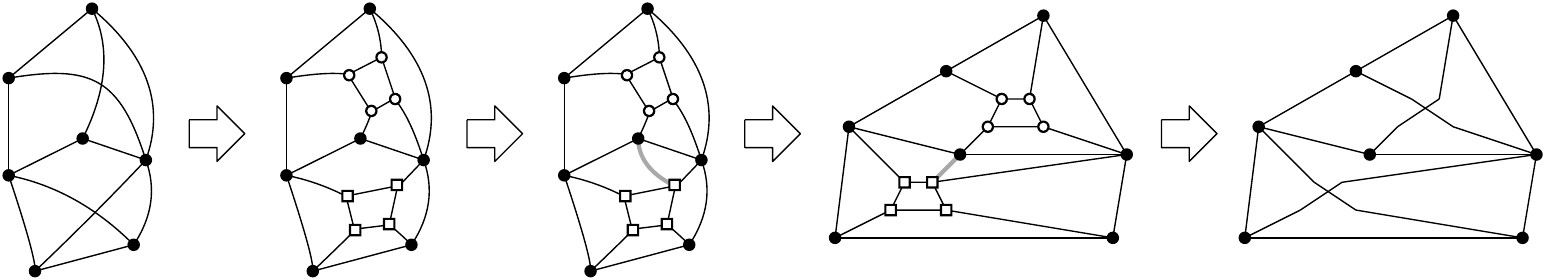}
\caption{Illustration of a technique for computing $2$-bend drawings of $1$-plane graphs in $O(n^4)$ area.\label{fi:1-planar-2bends}}
\end{figure}

As reported in Section~\ref{se:stretchability}, the class of $1$-plane graphs admitting a straight-line drawing has been characterized~\cite{th-rdg-88}. The same characterization has been re-discovered by Hong et al., who also show that there exist $1$-plane graphs such that every embedding-preserving straight-line drawing requires exponential area~\cite{DBLP:conf/cocoon/HongELP12}. Later on, it was proven that every $3$-connected $1$-planar graph admits a $1$-planar embedding that can be realized as a straight-line drawing except for at most one edge~\cite{DBLP:conf/gd/AlamBK13}; this drawing has quadratic area. On the other hand, it is immediate to see that every $1$-plane graph admits a $1$-bend drawing, as it suffices to replace each crossing with a dummy vertex and compute a straight-line drawing of the resulting plane graph (possible overlaps of bend points can be removed by slight perturbations). Furthermore, it is not difficult to see that every $1$-plane graph can be drawn with at most two bends per edge in $O(n^4)$ area. The idea is the following, refer to Fig.~\ref{fi:1-planar-2bends}. Replace each crossing with a dummy face of degree four; augment the resulting plane graph to be $3$-connected so that no edge is inserted in the dummy faces; compute a straight-line planar drawing such that all the faces are strictly convex and the area of the drawing is $O(n^2)\times O(n^2)$~\cite{DBLP:journals/corr/abs-cs-0507030}; replace all dummy vertices with bend points, remove all dummy edges, and reinsert the crossings. Chaplik, Lipp, Wolff, and Zink prove that every $n$-vertex $1$-plane graph actually admits an embedding-preserving $1$-bend RAC drawing; if two bends per edge are allowed, a RAC drawing can be computed in $O(n^6)$ area~\cite{DBLP:conf/ewcg/Chaplick18,zink-17}. They also prove that every NIC-plane graph admits an embedding-preserving $1$-bend RAC drawing in quadratic area. We remark that the existence of $1$-bend $1$-planar RAC drawings for every $1$-plane graph was already known~\cite{Bekos2017}, but only assuming that the drawing algorithm can change the $1$-planar embedding of the input graph. We also remark that for a simple family of $1$-plane graphs, called \emph{kite-triangulations}, an $\Omega(n^3)$ area lower bound for embedding-preserving straight-line RAC drawings is established~\cite{DBLP:journals/jgaa/AngeliniCDFBKS11}. A kite-triangulation is obtained by augmenting a plane triangulation with edges inside pairs of adjacent faces. Additional results concerned with other restricted classes of $1$-planar graphs, such as the IC-planar graphs, are surveyed in~\cite{DBLP:journals/csr/KobourovLM17}.

The edge complexity of RAC drawings has also been studied regardless of the number of crossings per edge. It is known that every $n$-vertex graph has a $3$-bend RAC drawing in $O(n^4)$ area~\cite{DBLP:journals/tcs/DidimoEL11}, while $1$-bend and $2$-bend RAC drawable graphs have at most $6.5n-13$ edges or $74.2n$ edges, respectively~\cite{DBLP:journals/comgeo/ArikushiFKMT12}. If we allow up to four bends per edge, every graph can be drawn RAC in $O(n^3)$ area~\cite{DBLP:journals/mst/GiacomoDLM11}. For graphs with bounded vertex degree, we have the following results~\cite{DBLP:journals/jgaa/AngeliniCDFBKS11}: $(i)$ every graph with vertex degree at most three has a $1$-bend RAC drawing; $(ii)$ every graph with vertex degree at most six has a $2$-bend RAC drawing. Both in $(i)$ and $(ii)$, the drawing is computed in $O(n)$ time and has $O(n^2)$ area.

Concerning ACE$\alpha$ drawings with one or two bends per edge, it is proven that they have at most $27n$ edges and $385n$ edges, respectively~\cite{aft-sgapd-10,DBLP:journals/siamdm/AckermanFT12}. On the other hand, every graph admits an ACL$\alpha$ drawing with one bend per edge in $O(n^2)$ area~\cite{DBLP:journals/mst/GiacomoDLM11}.

Table~\ref{ta:edge-complexity} summarizes the main results concerning trade-offs between the number of bends per edge and the area requirement of beyond-planar drawings.

\begin{table}[h!]%
\centering
\caption{Edge complexity and area requirement trade-offs. $\Delta$ denotes the maximum vertex degree. The question mark  indicates that no area bound is known.\label{ta:edge-complexity}}\smallskip{%
\renewcommand{\arraystretch}{1.3}
\footnotesize
%\resizebox{\textwidth}{!}{
\begin{tabular}{|l|c|c|c|c|c|l|}
\hline

{} & \multicolumn{5}{c|}{\textsc{Number of bends per edge}} & {}\\

\hline
 \textsc{Graph Family} & \textsc{0} & \textsc{1} & \textsc{2} & \textsc{3} & \textsc{4} & {\textsc{References}}\\\hline
 {IC-planar} & {$O(n^2)$} & {$O(n^2)$} & {$O(n^2)$} & {$O(n^2)$} & {$O(n^2)$}& {\cite{DBLP:journals/tcs/BrandenburgDEKL16}}\\\hline
 {NIC-plane RAC} & {?} & {$O(n^2)$} & {$O(n^2)$} & {$O(n^2)$} & {$O(n^2)$}& {\cite{zink-17}}\\\hline
 {$1$-planar} & {$\Omega(2^n)$} & {?} & {$O(n^4)$} & {$O(n^4)$} & {$O(n^4)$}& {\cite{DBLP:conf/cocoon/HongELP12}, this paper}\\\hline
 {1-plane RAC} & {$\Omega(2^n)$} & {?} & {$O(n^6)$} & {$O(n^6)$}& {$O(n^6)$}& {\cite{DBLP:conf/cocoon/HongELP12,zink-17}}\\\hline
 {RAC} & {$\Omega(n^2)$} & {$O(n^2)$, if $\Delta \le 6$} & {$O(n^2)$, if $\Delta \le 3$} & {$O(n^4)$}& {$O(n^3)$}& {\cite{DBLP:journals/jgaa/AngeliniCDFBKS11,DBLP:journals/mst/GiacomoDLM11,DBLP:journals/tcs/DidimoEL11}}\\\hline
 {ACL$\alpha$} & {?} & {$O(n^2)$} & {$O(n^2)$} & {$O(n^2)$}& {$O(n^2)$}& {\cite{DBLP:journals/mst/GiacomoDLM11}}\\\hline
\end{tabular}
%}
}
%\begin{tabnote}%
%\Note{Notes.}{$\Delta$ denotes the maximum vertex degree. The question mark  indicates that no area bound is known.}
%\end{tabnote}
\end{table}%

\subsection{Vertex Complexity}\label{ss:Vertex-Coplexity}

A \emph{bar visibility representation} of a graph $G$ maps each vertex of $G$ to a distinct horizontal segment, called \emph{bar}, and each edge of $G$  to a vertical unobstructed segment, called \emph{visibility}, between the two bars corresponding to its end-vertices. This kind of representation is intrinsically planar, and on the other hand every planar graph admits this representation\footnote{Here we refer to the so-called \emph{weak model}, in which the existence of a visibility between a pair of bars does not necessarily imply the existence of an edge in the graph between the two corresponding vertices.} (see, e.g.,~\cite{TamassiaTollis86,Wismath85}). In order to realize non-planar graphs, other visibility models have been proposed, such as bar $k$-visibility representations and rectangle visibility representations.

In a \emph{bar $k$-visibility representation} each visibility intersects at most $k$ bars~\cite{DBLP:journals/jgaa/DeanEGLST07}. In particular, it is proven that every $1$-planar graph has a bar $1$-visibility representation~\cite{BrandenburgJGAA14,Evans0LMW14}.
A \emph{rectangle visibility representation} maps each vertex to an axis-aligned rectangle and each edge to a horizontal or vertical visibility between the two corresponding rectangles~\cite{Shermer96}. In general there are $1$-plane graphs, and even IC-plane graphs, that do not admit such a representation. Biedl et al. describe a linear-time algorithm to test whether a $1$-plane graph admits an embedding-preserving rectangle visibility representation, and to compute one if it exists~\cite{DBLP:conf/compgeom/BiedlLM16}. The algorithm  is based on a characterization that extends the set of obstructions used by Thomassen to characterize the stretchable $1$-plane graphs~\cite{th-rdg-88} (see Section~\ref{se:stretchability}). The rectangle visibility model has been recently generalized to extend the class of representable graphs; namely, in a generalized model, called \emph{ortho-polygon visibility representations}, the vertices are drawn as general orthogonal polygons~\cite{DiGiacomo2017}. Di Giacomo et al. prove that every $1$-plane graph admits an embedding-preserving ortho-polygon visibility representation~\cite{DiGiacomo2017}. If the input graph is $3$-connected, one can construct a representation such that each vertex has at most $12$ reflex corners, while $2$ reflex corners may be needed. If the graph is only $2$-connected, it may require at least one vertex with a linear number of reflex corners.

\newcommand{\SA}{\includegraphics{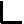}~}
\newcommand{\SB}{\includegraphics{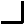}~}
\newcommand{\SC}{\includegraphics{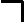}~}
\newcommand{\SD}{\includegraphics{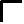}~}
\newcommand{\shapes}{$\{\SA,\SB,\SC,\SD\}$}

A further visibility model, called \emph{L-visibility representations}, maps every vertex to a horizontal and a vertical segment sharing an endpoint (i.e., by an L-shape in the set \shapes), and each edge to a horizontal or vertical visibility segment joining the two L-shapes corresponding to its two end-vertices. It is proven that every IC-planar graph admits an L-visibility representation~\cite{LiottaM16}.

Visibility representations of $1$-planar graphs have been studied also in 3D. A \emph{$z$-parallel visibility representation} maps the vertices of a graph to isothetic disjoint rectangles parallel to the $xy$-plane, and the edges to visibilities parallel to the $z$-axis. Every $1$-plane graph has a $z$-parallel visibility representation~\cite{DBLP:conf/gd/Angelini17}. The computed drawing is such that there is a plane orthogonal to the rectangles of the representation and the intersection of this plane with the representation defines a bar $1$-visibility representation of the graph.

Finally, Alam et al. studied contact representations in which vertices are represented by axis-aligned polyhedra in 3D and edges are realized by non-zero area common boundaries between corresponding polyhedra~\cite{AlamEKPTU15}. They prove that every optimal $1$-plane graph can be realized as a contact representation where vertices are axis-aligned boxes if it contains no separating $4$-cycles, or as a contact representation where vertices are L-shaped polyhedra otherwise.

Table~\ref{ta:vertex-complexity} summarizes the main results on visibility and contact representations of $1$-planar graphs.

\begin{table}[h!]%
\caption{Vertex complexity. $\times$ indicates that there are instances of the graph family not admitting that visibility or contact representation. \checkmark means that all instances in the graph family can be drawn with that visibility or contact representation. The question mark means that it not known whether the answer is $\times$ or \checkmark. LEGEND: RVR = rectangle visibility representation, B$1$VR = bar $1$-visibility representation, $k$OPVR = ortho-polygon visibility representation, LVR = L-visibility representation, ZPR = $z$-parallel visibility representation, BCR = box contact representation, LCR = L-shaped contact representation.\label{ta:vertex-complexity}}\smallskip{%
\renewcommand{\arraystretch}{1.3}
\footnotesize
\resizebox{\textwidth}{!}{
\begin{tabular}{|l|c|c|c|c|c|c|c|c|l|}
\hline

{} & \multicolumn{5}{c|}{\textsc{Visibility}} & \multicolumn{2}{c|}{\textsc{Contact}} & {}\\

\hline
 \textsc{Graph Family} & \textsc{RVR} & \textsc{B$1$VR} & \textsc{OPVR} & \textsc{LVR} & \textsc{ZPR} & \textsc{BCR}& \textsc{LCR} & {\textsc{References}}\\\hline
{IC-planar} & {\checkmark} & {\checkmark} & {\checkmark} &{\checkmark} & {\checkmark} & {?} & {?}& \multirow{4}{*}{\shortstack{\cite{AlamEKPTU15} \cite{DBLP:conf/gd/Angelini17}\\ \cite{DBLP:conf/compgeom/BiedlLM16} \cite{BrandenburgJGAA14}\\\cite{DiGiacomo2017} \cite{Evans0LMW14}\\ \cite{LiottaM16}}}\\
{optimal $1$-planar with no separating $4$-cycles} & {$\times$} & {\checkmark} & {\checkmark} & {\checkmark} & {\checkmark} & {\checkmark} & {\checkmark}& {}\\
{optimal $1$-planar} & {$\times$} & {\checkmark} & {\checkmark} & {\checkmark} & {\checkmark} & {?} & {\checkmark}& {}\\
{$1$-planar} & {$\times$} & {\checkmark} & {\checkmark} & {$\times$} & {\checkmark} & {?} & {?}& {}\\\hline
\end{tabular}
}
}
%\begin{tabnote}%
%\Note{Legend.}{RVR = rectangle visibility representation, B$1$VR = bar $1$-visibility representation, $k$OPVR = ortho-polygon visibility representation, LVR = L-visibility representation, ZPR = $z$-parallel visibility representation, BCR = box contact representation, LCR = L-shaped contact representation.}
%\Note{Notes.}{$\times$ indicates that there are instances of the graph family not admitting that visibility or contact representation. \checkmark means that all instances in the graph family can be drawn with that visibility or contact representation. The question mark means that it not known whether the answer is $\times$ or \checkmark.}
%\end{tabnote}
\end{table}%

%\subsection{Area}\label{ss:Area}
%\subsection{Area-crossings Trade-offs for Planar graphs}
\subsection{Area-Crossing Trade-offs for Planar Graphs}\label{ss:tradeoffs-planar}
Many papers in graph drawing study the area requirement of planar straight-line drawings.
Different popular results establish that an $n$-vertex planar graph can always be drawn with straight-line edges in $O(n^2)$ area~\cite{DBLP:journals/combinatorica/FraysseixPP90,DBLP:conf/soda/Schnyder90}.
It is also proven that this bound is worst-case optimal for the family of planar graphs, as there are infinitely many planar graphs that require quadratic area to be drawn in the plane without edge crossings~\cite{DBLP:journals/combinatorica/FraysseixPP90}.
Several attempts have been done to prove the existence of straight-line
planar drawings with $o(n^2)$ area for specific sub-families of planar graphs, such as trees, outerplanar graphs, and series-parallel graphs (see, e.g.,~\cite{DBLP:journals/comgeo/GiacomoDLM13} for references on these results).

A natural question that arises from the aforementioned results is whether allowing some edge crossings may help to reduce the area of a drawing of a planar graph. In other words, what is the area requirement of beyond-planar drawings of planar graphs?
Wood shows that, for any fixed positive integer $k > 0$, all $k$-colorable graphs have a straight-line drawing in linear area; this implies that planar graphs always admit $O(n)$ straight-line drawings with crossing edges~\cite{DBLP:journals/comgeo/Wood05}. However, the technique by Wood can give rise to drawings where some edges contain a linear number of crossings and the angles at which two edges cross can be arbitrarily small.

The use of edge crossings that form large angles is studied in different papers. Straight-line RAC drawings of planar graphs may require $\Omega(n^2)$ area~\cite{DBLP:journals/jgaa/AngeliniCDFBKS11}, thus right angle crossing drawings do not help to reduce the area requirement bound for planar graphs in the general case. On the positive side, for infinitely many values of $n$, there exists an $n$-vertex planar graph whose requirement is $\Theta(n^2)$ for straight-line planar drawings and $\Theta(n)$ for straight-line RAC drawings~\cite{DBLP:conf/gd/Kreveld10}. Analogous results hold for other aesthetics such as uniform edge length and angular resolution~\cite{DBLP:conf/gd/Kreveld10}. In addition, every planar graph admits an ACL$\alpha$ drawing with two bends per edge in $O(n^\frac{5}{3})$ area~\cite{DBLP:conf/s-egc/AngeliniBDFHKLL11} and every planar graph with vertex-degree at most $\Delta$ admits a $4$-bend RAC drawing in $O(n \sqrt{\Delta n})$ area~\cite{DBLP:conf/s-egc/AngeliniBDFHKLL11}. Note that, if $\Delta$ is a sublinear function of $n$, these RAC drawings have subquadratic area.

Compact non-planar drawings of planar graphs with constant or sublinear number of crossings per edge are also studied~\cite{DBLP:journals/comgeo/GiacomoDLM13}.
Every $n$-vertex outerplanar graph admits a straight-line drawing with $O(\frac{n}{\log n})$ crossings per edge
in $O(n \log n)$ area. Also, for any given $\epsilon > 0$, every $n$-vertex outerplanar graph admits a straight-line drawing with $O(n^{1-\epsilon})$ crossings per edge in $O(n^{1+\epsilon})$ area, which gives a clear trade-off scheme between area requirement and number of crossings per edge. Both these results are based on a linear-time drawing algorithm, which can also be applied to other sub-families of planar graphs that admit a ``level'' drawing with specific properties, such as flat series-parallel graphs with bounded degree (see~\cite{DBLP:journals/comgeo/GiacomoDLM13} for a definition of these families).
On the other hand, if we insist in having a constant number of crossings per edge, planar and non-planar drawings of planar graphs have in general the same area requirement. This is true even for series-parallel graphs~\cite{DBLP:journals/cj/GiacomoDLM17}.

On the positive side, every $n$-vertex planar graph has a straight-line $o(n)$-quasi planar drawing in $o(n^2)$ area. More precisely, 
by combining drawing techniques in~\cite{DBLP:journals/cj/GiacomoDLM17} with results on the track number of planar graphs~\cite{DBLP:journals/jct/Dujmovic15}, one can prove that every $n$-vertex planar graph admits either a straight-line $O(\log n)$-quasi planar drawing in  $O(n \log^{3} n)$ area or a straight-line $O(\log^{2} n)$-quasi planar drawing in $O(n \log n)$ area.
%every $n$-vertex planar graph admits a straight-line drawing with either $O(n \log^{48} n)$ area and $O(\log^{16} n)$ thickness or $O(n \log^{16} n)$ area and $O(\log^{32} n)$ thickness (the thickness of a drawing is the minimum number of colors that can be assigned to the edges so that each color class induces a planar drawing)~\cite{DBLP:journals/cj/GiacomoDLM17}. Since a drawing with thickness $h$ is $(h+1)$-quasi planar, the result implies that every $n$-vertex planar graph has a straight-line $o(n)$-quasi planar drawing in $o(n^2)$ area. The proof of the aforementioned area-thickness bounds depends on known bounds on the track number of planar graphs. An improved bound on the track number of planar graphs~\cite{DBLP:journals/jct/Dujmovic15} can be directly applied to that proof to obtain either $O(n \log^{3} n)$ area and $O(\log n)$ thickness or $O(n \log n)$ area and $O(\log^{2} n)$ thickness.
Also, every partial $2$-tree admits a linear-area straight-line drawing with thickness at most $10$, and hence a linear-area $11$-quasi planar straight-line drawing~\cite{DBLP:journals/cj/GiacomoDLM17}. Furthermore, outerplanar graphs and flat series-parallel graphs with bounded vertex degree (which are partial $2$-trees) have quasi planar and $5$-quasi planar drawings in linear area, respectively~\cite{DBLP:conf/wg/GiacomoDLM12}.

Table~\ref{ta:area-crossings} summarizes the main results concerning area-crossings trade-offs for beyond-planar drawings of planar graphs.

\begin{table}[h!]%
\caption{Area-Crossings Trade-Offs for Planar Graphs. $\Delta$ denotes the maximum vertex degree. The question mark indicates that bounds better than those for planar drawings are not known. For reasons of space, the table does not report the corresponding references.\label{ta:area-crossings}}\smallskip{%
\renewcommand{\arraystretch}{1.8}
\tiny
\resizebox{\textwidth}{!}{
\begin{tabular}{|l|c|c|c|c|c|c|c|c|c|}
\hline
 \multirow{2}{*}{\textsc{
\diagbox{\textsc{Planar}}{\textsc{Beyond-planar}}}} & \multicolumn{2}{c|}{\textsc{straight-l. $k$-planar}} & \multicolumn{2}{c|}{\textsc{straight-l. $k$-quasi planar}} & \multicolumn{2}{c|}{\textsc{$k$-bend RAC}} & \multicolumn{2}{c|}{\textsc{$k$-bend ACL$\alpha$}}\\
& {\textsc{Area}} &  {$k$} &  {\textsc{Area}} &  {$k$} & {\textsc{Area}} &  {\textsc{$k$}} & {\textsc{Area}} &  {\textsc{$k$}}\\\hline
\multirow{2}{*}{outerplanar} & {$O(n^{1+\epsilon})$} & {$O(n^{1-\epsilon})$} & \multirow{2}{*}{$O(n)$} & \multirow{2}{*}{$3$} & \multirow{2}{*}{?} & \multirow{2}{*}{} & \multirow{2}{*}{?} & \multirow{2}{*}{} \\
& {$O(n \log n)$} & {$O(\frac{n}{\log n})$} & {} & {} & {} & {} & {} & {} \\\hline
{flat series-parallel bounded degree} & {$O(n^{1+\epsilon})$} & {$O(n^{1-\epsilon})$} & \multirow{2}{*}{$O(n)$} & \multirow{2}{*}{$5$} & \multirow{2}{*}{?} & \multirow{2}{*}{} & \multirow{2}{*}{?} & \multirow{2}{*}{} \\
& {$O(n \log n)$} & {$O(\frac{n}{\log n})$} & {} & {} & {} & {} & {} & {} \\\hline
{partial $2$-tree} & {$\Omega(n2^{\sqrt{\log n}})$} & {$O(1)$} & {$O(n)$} & {$11$} & {?} & {} & {?} & {}\\\hline
\multirow{2}{*}{planar} & {$\Omega(n^2)$} & {$O(1)$} & {$O(n \log^3 n)$} & {$O(\log n)$} & {$\Omega(n^2)$} & {$0$} & \multirow{2}{*}{$O(n^\frac{5}{3})$} & \multirow{2}{*}{$2$}\\
& {$O(n)$} & {$O(n)$} & {$O(n \log n)$} & {$O(\log^2 n)$} & {$O(n \sqrt{\Delta n})$} & {$4$} & {} & {}
\\\hline
\end{tabular}
}
}
%\begin{tabnote}%
%\Note{Notes.}{$\Delta$ denotes the maximum vertex degree. The question mark indicates that bounds better than those for planar drawings are not known. For reasons of space, the table does not report the corresponding references.}
%\end{tabnote}
\end{table}%

%\subsection{Open Problems}
\myparagraph{Open Problems.} Several open problems can be deduced by looking at the ``?'' in Tables~\ref{ta:edge-complexity}, \ref{ta:vertex-complexity}, and \ref{ta:area-crossings}. For example:

\begin{problem}\label{op:edge-complexity}
It is known that every $1$-planar graph has a RAC drawing with at most 1-bend per edge~\cite{Bekos2017}. However, it is not known whether such a drawing can be computed in polynomial area.
\end{problem}

About the problem above, it has been recently shown that NIC-plane graphs admit an embedding-preserving $1$-bend RAC drawing in quadratic area~\cite{DBLP:conf/ewcg/Chaplick18}.

\begin{problem}\label{op:vertex-complexity}
Do $1$-planar graphs admit a box contact representation? The question is interesting even for subfamilies of $1$-planar graphs, such as IC-planar graphs.
\end{problem}

\begin{problem}\label{op:area}
Do planar graphs admit a $k$-bend RAC drawing in subquadratic area with $k \geq 4$? 
\end{problem}

When $k=4$, the answer is affirmative if the maximum vertex-degree is sublinear~\cite{DBLP:conf/s-egc/AngeliniBDFHKLL11}.
\section{Constraints}\label{se:constraints}
Several papers concentrate on beyond-planar drawings where the vertices and/or the edges have additional geometric constraints. We discuss the main scenarios studied in the literature.

\subsection{Vertices on lines, circles, and external boundary}\label{sse:constraints-lines-circles}
The study of graph layouts in which vertices are placed on a given set of horizontal lines, often called \emph{layers}, or on a set of concentric circles, has a well-established tradition in graph drawing~\cite{dett-gd-99,DBLP:reference/crc/GiacomoDL13}.

\myparagraph{2-layer drawings.} In the beyond planarity context, straight-line $2$-layer drawings have been investigated for RAC, $1$-planar, and fan-planar graphs,  in terms of both edge density and recognition. In a $2$-layer drawing the vertices are distributed along two horizontal layers, and vertices of the same layer cannot be adjacent. Thus, the graph is necessarily bipartite. The study of $2$-layer drawings has two main motivations: $(i)$ they are a natural way to visually convey bipartite graphs; $(ii)$ algorithms that compute $2$-layer drawings are building blocks of the popular Sugiyama's framework~\cite{DBLP:journals/tsmc/SugiyamaTT81}, used to draw graphs on multiple horizontal layers.

\begin{figure}[tb]
\centering
\subfigure[]{\includegraphics[width=0.2\columnwidth,page=1]{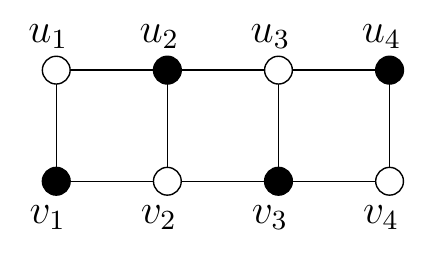}\label{fi:ladders-1}}\hfil
\subfigure[]{\includegraphics[width=0.2\columnwidth,page=2]{figs/ladders}\label{fi:ladders-2}}\hfil
\subfigure[]{\includegraphics[width=0.2\columnwidth,page=3]{figs/ladders}\label{fi:ladders-3}}
\caption{(a) A ladder $G$; (b) A $2$-layer $1$-planar drawing of $G$ that is also RAC. (c) A snake.\label{fi:ladders}}
\end{figure}

Characterizations for those graphs that admit either $2$-layer RAC drawings~\cite{DBLP:journals/algorithmica/GiacomoDEL14}, or $2$-layer $1$-planar drawings~\cite{D13}, or $2$-layer fan-planar drawings~\cite{DBLP:journals/jgaa/BinucciCDGKKMT17} are known; all of them can be regarded as generalizations of caterpillars, the class of $2$-layer planar drawable graphs~\cite{emw-ecp-86}. The upper bounds on the edge density of these graph families are reported in Table~\ref{ta:density-constrained}. In particular, the optimal $2$-layer $1$-planar graphs coincide with the optimal $2$-layer RAC graphs, and have $1.5n - 2$ edges, where $n$ is the number of vertices of the graph. They are called \emph{ladders} and consist of two paths of the same length $\langle u_1,u_2, \dots , u_\frac{n}{2} \rangle$ and $\langle v_1,v_2, \dots , v_\frac{n}{2} \rangle$, plus the edges $(u_i, v_i)$, $(i = 1, 2, \dots , \frac{n}{2})$; see Figs.~\ref{fi:ladders-1} and~\ref{fi:ladders-2}. A maximal $2$-layer fan-planar graph is called \emph{snake}, and is obtained from an outerplane ladder, by adding, inside each internal face, an arbitrary number (possibly none) of paths of length two joining a pair of non-adjacent vertices of the face. Intuitively, a snake is a bipartite planar graph composed of a chain of complete bipartite graphs $K_{2,h}$; see Fig.~\ref{fi:ladders-3}. The family of optimal $2$-layer fan-planar graphs on $n$ vertices includes $K_{2, n-2}$~\cite{DBLP:journals/tcs/BinucciGDMPST15}.

From the algorithmic point of view, there exist linear-time testing and drawing algorithms for $2$-layer RAC graphs~\cite{DBLP:journals/algorithmica/GiacomoDEL14} and for $2$-connected $2$-layer fan-planar  graphs~\cite{DBLP:journals/jgaa/BinucciCDGKKMT17}, while it is still open the complexity of recognizing $2$-layer fan-planar graphs when the graph is only $1$-connected. For $2$-layer RAC graphs, it is also possible to compute, in $O(n^2\log n)$ time, a $2$-layer RAC drawing with minimum number of crossings~\cite{DBLP:journals/algorithmica/GiacomoDEL14}. We remark that the crossing minimization problem is a well-known \npc problem in the general case, which remains hard even for $2$-layer drawings without the restriction that crossing edges are orthogonal~\cite{DBLP:journals/algorithmica/EadesW94}.

\myparagraph{$k$-page Drawings.} Let $\ell$ be a line, called \emph{spine}. In a \emph{$k$-page drawing}\footnote{We recall that $k$-page drawings are among the oldest and more common graph drawing conventions and they can be found with different names in the literature. For example, they are sometimes called \emph{linear layouts} (see, e.g.,~\cite{DBLP:journals/dmtcs/DujmovicW04}); also, a $2$-page drawing is sometimes called {\em permutation representation} (see, e.g.,~\cite{5249682}) or {\em linear drawing} (see, e.g.,~\cite{DBLP:reference/crc/BuchheimCGJM13}), while a $1$-page drawing is also called {\em arc diagram} (see, e.g.,~\cite{DBLP:conf/infovis/Wattenberg02})} of a graph each vertex is mapped to a distinct point of  $\ell$ and each edge is drawn as a semicircle in one of $k$ distinct half-planes incident to $\ell$. Each of these half-planes is called a {\em page}. Binucci et al. study the problem of computing $2$-page and $1$-page drawings where the number of crossings per edge does not depend on the size of the input graph~\cite{DBLP:journals/ejc/BinucciGHL18} . They describe linear-time algorithms to compute $2$-page drawings of planar $3$-trees with at most $2\Delta$ crossings per edge, where $\Delta$ is the maximum vertex-degree of the graph, and $1$-page drawings of partial $2$-trees with at most $\Delta^2$ crossings per edge. In both cases, the authors show that the number of crossings per edge cannot be bounded by a constant.

If no two edges cross, a $k$-page drawing is called \emph{$k$-page book embedding} (or \emph{$k$-stack layout}). The minimum value of $k$ such that a graph $G$ has a $k$-page book embedding is the \emph{book thickness} (or \emph{stack number}) of $G$. An $O(\log n)$ upper bound on the  book thickness of $k$-planar graphs has been recently proved~\cite{JGAA-454}, while for $1$-planar graphs $O(1)$ upper bounds are knwon~\cite{DBLP:journals/corr/AlamBK15,DBLP:journals/algorithmica/BekosBKR17}.

\myparagraph{Outer and convex drawings.} Beyond-planar graphs that admit a layout in which all vertices lie on a circle enclosing the whole drawing or, more in general, on the external boundary of the drawing, can be regarded as generalizations of the outerplanar graphs. For any integer $k > 0$, a $k$-planar drawing with this property is an \emph{outer $k$-planar} graph. Outer $1$-planar graphs have at most $2.5n-4$ edges~\cite{AuerBBGHNR16,D13}, which is a tight bound. They can be recognized in linear time~\cite{AuerBBGHNR16,DBLP:journals/algorithmica/HongEKLSS15} and drawn in linear time, both as straight-line drawings with all vertices on the external face in $O(n^2)$ area and as visibility representations in $O(n \log n)$ area~\cite{AuerBBGHNR16}. Hong and Nagamochi extended the recognition problem to \emph{full} outer $2$-planar graphs, i.e., outer $2$-planar graphs with no crossing on the external boundary~\cite{DBLP:conf/wg/HongN15}. They prove that also this class of graphs can be recognized in linear time.

Density and recognition results are described also for \emph{outer fan-planar} graphs. They have at most $3n-5$ edges (see Table~\ref{ta:density-constrained}), but, in contrast with outer $1$-planar graphs, the complexity of the recognition problem is not known in general, while it is linear-time solvable for \emph{maximal} outer fan-planar graphs~\cite{bcghk-16}.

\begin{table}[bt]%
\centering
\caption{Density for Constrained Families.\label{ta:density-constrained}}\smallskip{%
\renewcommand{\arraystretch}{1.3}
\footnotesize
%\resizebox{\textwidth}{!}
%{
\begin{tabular}{|l|c|c|l|}
\hline
 \textsc{Graph Family} & \textsc{Max. Num. Edges} & \textsc{Tight} & \textsc{References} \\\hline
 {$2$-layer $1$-planar} & {$1.5n-2$} & {$\checkmark$} & {\cite{D13}}  \\\hline
 {$2$-layer RAC} & {$1.5n-2$} & {$\checkmark$} & {\cite{DBLP:journals/algorithmica/GiacomoDEL14}}  \\\hline
 {$2$-layer fan-planar} & {$2n-4$} & {$\checkmark$} & {\cite{DBLP:journals/tcs/BinucciGDMPST15}}  \\\hline
 {outer $1$-planar} & {$2.5n-4$} & {$\checkmark$} & {\cite{D13,AuerBBGHNR16}}  \\\hline
 {outer fan-planar} & {$3n-5$} & {$\checkmark$} & {\cite{DBLP:journals/tcs/BinucciGDMPST15}}  \\\hline
 {convex geometric $k$-quasi planar} & {$2(k-1)n-\binom{2k-1}{2}$} & {$\checkmark$} & {\cite{DBLP:journals/jct/CapoyleasP92}} \\\hline
 {convex geometric fan-crossing-free} & {$\lfloor 5n/2 - 4 \rfloor$} & {$\checkmark$} & {\cite{bkv-03}} \\\hline

\end{tabular}
%}
}
\end{table}%

Table~\ref{ta:density-constrained} also reports some other tight bounds for geometric beyond-planar graphs with all vertices in convex position, such as $k$-quasi planar graphs and fan-crossing-free graphs. Bounds for more specific families are listed in the geometric graph theory chapter of the Handbook of Discrete and Computational Geometry~\cite{DBLP:reference/cg/Pach04a}. Note that, the edge crossings in a geometric graph with all vertices in convex position are the same as the edge crossings of a $1$-page drawing in which the order of the vertices along the spine is the same as the circular order of the vertices on the convex polygon.

\begin{table}[h!]%
\centering
\caption{Recognition for Constrained Families.\label{ta:recognition-constrained}}\smallskip{%
\renewcommand{\arraystretch}{1.3}
\footnotesize
%\resizebox{\textwidth}{!}
%{
\begin{tabular}{|l|c|l|}
\hline
 \textsc{Graph Family} & \textsc{Complexity} & \textsc{References} \\\hline
 {$2$-layer RAC} & {$O(n)$} & {\cite{DBLP:journals/algorithmica/GiacomoDEL14}}\\\hline
 {$2$-connected $2$-layer fan-planar} & {$O(n)$} & {\cite{DBLP:journals/jgaa/BinucciCDGKKMT17}}\\\hline
 {outer $1$-planar} & {$O(n)$} & {\cite{AuerBBGHNR16,DBLP:journals/algorithmica/HongEKLSS15}}\\\hline
 {full outer $2$-planar} & {$O(n)$} & {\cite{DBLP:conf/wg/HongN15}}\\\hline
 {maximal outer fan-planar} & {$O(n)$} & {\cite{bcghk-16}}\\\hline
 {circular RAC} & {$O(n)$} & {\cite{DBLP:journals/tcs/DehkordiEHN16}}\\\hline
 {upward RAC} & {\nph} & {\cite{DBLP:journals/jgaa/AngeliniCDFBKS11}}\\\hline
 {simultaneous RAC} & {\nph} & {\cite{DBLP:journals/jgaa/Grilli17}}\\\hline
 {$k$-SEFE} & {\npc} & {\cite{DBLP:journals/jgaa/Grilli17}}\\\hline
 {straight-line point-set RAC} & {\nph} & {\cite{DBLP:conf/walcom/FinkHMSW12}}\\\hline
\end{tabular}
%}
}
\end{table}%

Several interesting properties of convex geometric $k$-planar and $k$-quasi planar graphs have been recently investigated~\cite{DBLP:journals/corr/abs-1708-08723}. It is shown that convex geometric $k$-planar graphs are $\lfloor \sqrt{4k+1} \rfloor +1$-degenerate and consequently $\lfloor \sqrt{4k+1} \rfloor + 2$-colorable. Furthermore, they have a balanced separator of size at most $2k+3$, which allows the design of a quasi-polynomial time recognition algorithm for this class of graphs, i.e., the recognition problem is not NP-hard unless ETH (Exponential Time Hypothesis~\cite{DBLP:journals/jcss/ImpagliazzoP01}) fails. Convex geometric $k$-planar graphs in which all the vertices form a simple cycle can be recognized in linear time, because they can be expressed in extended monadic second-order logic and have bounded threewidth. In~\cite{DBLP:journals/corr/abs-1708-08723} we can also find families of $3$-quasi planar graphs that are convex geometric and families that are not convex geometric. 

\smallskip
A characterization of the class of graphs that admit a RAC drawing in which all vertices lie on a circle, along with a linear-time testing and layout algorithm are given in~\cite{DBLP:journals/tcs/DehkordiEHN16}. 
%For those graphs that do not have a circular RAC drawing, the authors describe a practical quadratic programming approach for increasing crossing angles in a circular drawing.
Outer $1$-planar graphs have been studied in combination with other types of constraints or beyond-planar graphs. Dehkordi and Eades show that every outer $1$-planar graph admits an outer RAC drawing~\cite{DBLP:journals/ijcga/DehkordiE12}. 
Di Giacomo et al. study how to draw outer $1$-planar graphs with a small number of edge slopes~\cite{DBLP:journals/jgaa/GiacomoLM15}. They prove that every outer $1$-planar graph admits an outer $1$-planar straight-line drawing that uses $O(\Delta)$ different slopes, where $\Delta$ is the maximum vertex-degree of the graphs.

\subsection{Upward RAC Drawings, Simultaneous RAC and Point-set RAC Embedding}\label{sse:constraints-rac}

RAC drawings have been studied in combination with other popular graph drawing conventions that impose additional geometric constrains on the layout, namely the \emph{upward drawing} convention for directed graphs (digraphs for short), the \emph{simultaneous embedding} convention for multiple graphs with the same vertex set, and the \emph{point-set embedding} convention. We discuss them below.

\myparagraph{Upward RAC Drawings.} An upward drawing of a digraph $G$ is such that each edge is drawn as a curve monotonically increasing in the vertical direction (see, e.g.,~\cite{dett-gd-99}). This kind of drawings is used to visually convey the structure of acyclic digraphs; extensions of upward drawings are also effective for the visualization of several types of networks, such as Petri Nets or social networks with both directed and undirected edges~\cite{DBLP:journals/algorithmica/BertolazziBD02,DBLP:journals/tcs/BinucciDP14}.
One of the most studied problems concerned with upward drawability is the design of algorithms that test whether a given planar digraph admits an upward planar drawing (i.e., an upward drawing without crossings), and in the affirmative case that compute one (refer to~\cite{DBLP:reference/algo/Didimo16} for a survey). Although polynomial-time testing algorithms are known for specific subfamilies of planar digraphs or when the digraph has a fixed embedding, the testing problem is \npc in the general~\cite{DBLP:journals/siamcomp/GargT01}. Also, although a planar digraph has an upward planar drawing if and only if it has a straight-line upward planar drawing~\cite{DBLP:journals/tcs/BattistaT88}, there are digraphs for which every straight-line upward planar drawing requires exponential area~\cite{DBLP:journals/tcs/BattistaT88,DBLP:journals/dcg/BattistaTT92}, while quadratic area is always achievable if we allow edge bends.

The use of right angle crossings for the upward drawability of acyclic planar digraphs has been investigated to overcome the aforementioned limits of upward planar drawings. In particular, it is natural to ask whether every planar acyclic digraph admits an upward RAC drawing and if every digraph with an upward RAC drawing admits one with straight-line edges in polynomial area. Unfortunately, both these questions have a negative answer and testing whether a planar digraph admits an upward RAC drawing remains \nph~\cite{DBLP:journals/jgaa/AngeliniCDFBKS11}.

\myparagraph{Simultaneous RAC embedding.} Given two planar graphs $G_1=(V,E_1)$ and $G_2=(V,E_2)$ with the same vertex set, a \emph{simultaneous embedding} of $G_1$ and $G_2$ is a pair of planar drawings, $\Gamma_1$ of $G_1$ and $\Gamma_2$ of $G_2$, such that each vertex $v \in V$ has the same position in $\Gamma_1$ and $\Gamma_2$ (the edges of $\Gamma_1$ are allowed to cross those of $\Gamma_2$). The simultaneous embedding problem is motivated by several practical scenarios. For example, they are useful to visually compare two social networks consisting of the same set of subjects but representing different types of relationships (e.g., friendships and work collaborations). It can also be applied to the visualization of an evolving network, whose edges change over time.
The simultaneous embedding problem was introduced a decade ago~\cite{DBLP:journals/comgeo/BrassCDEEIKLM07}, and since then it has been widely investigated in the literature (see, e.g.,~\cite{DBLP:reference/crc/BlasiusKR13} for a survey). When the edges are drawn as polylines, an important variant of the problem is the so called \emph{SEFE} (\emph{Simultaneous Embedding with Fixed Edges}), in which edges that occur in both graphs are drawn in the same way in both graphs (thus, they cannot be crossed).

A simultaneous RAC embedding of two graphs $G_1$ and $G_2$ has the additional property that the union of the two planar drawings $\Gamma_1$ and $\Gamma_2$ is a RAC drawing. The idea is to avoid sharp angles that may affect the readability of the whole layout. The simultaneous RAC embedding problem was originally introduced for straight-line drawings~\cite{DBLP:journals/jgaa/ArgyriouBKS13}, showing that only restricted pairs of planar subgraphs admit such an embedding; for example, a wheel and a cycle might not admit a straight-line simultaneous RAC embedding. The more general problem of deciding whether a pair of graphs admits a simultaneous RAC embedding (with straight-line edges) is \nph~\cite{DBLP:journals/jgaa/Grilli17}.
More recently, it has been proven that, if we use polyline edges, every pair of planar graphs has a simultaneous RAC embedding with at most six bends per edge~\cite{DBLP:journals/jgaa/BekosDKW16}. The number of bends per edge can be further reduced for some combinations of specific subfamilies of planar graphs, also in the SEFE scenario. For example, every pair consisting of a tree and of a matching has a simultaneously RAC embedding with fixed edges in which the tree is drawn with at most one bend per edge and the matching is drawn without bends. Moreover, simultaneous RAC embeddings with at most two bends per edge are possible for those pairs of graphs admitting a special type of visibility representation in which vertices are drawn as axis-aligned L-shapes and edges are vertical or horizontal lines of sight connecting pairs of such shapes~\cite{DBLP:journals/tcs/EvansLM16}.

The $k$-SEFE problem is a variant of SEFE, in which those edges that belong to one graph only may receive at most $k$ crossings each, where $k$ is a prescribed positive integer. Deciding if a pair of graphs is a positive instance of the $k$-SEFE problem is \npc for any fixed positive $k$~\cite{DBLP:journals/jgaa/Grilli17}.

We remark that another variant of simultaneous embedding in graph drawing beyond planarity considers a simultaneous embedding of two graphs $G_1$ and $G_2$ in which each of the two drawings $\Gamma_1$ and $\Gamma_2$ is allowed to have some desired type of crossings. This setting generalizes the classical simultaneous embedding instead of restricting it. For example, Di Giacomo et al. study \emph{simultaneous geometric quasi planar embedding (SGQPE)}, where each $\Gamma_i$ is a straight-line quasi-planar drawing~\cite{DBLP:journals/cj/GiacomoDLMW15}. They show for instance that a tree and a path always admit an SGQPE, in contrast with the negative result in the simultaneous geometric planar embedding setting~\cite{DBLP:journals/jgaa/AngeliniGKN12}. More in general, they prove that trees and other meaningful subfamilies of the outerplanar graphs admit an SGQPE. 

\myparagraph{Point-set RAC Embeddings.} Given a set $S$ of points in the plane and a graph $G$ with $n=|S|$ vertices, a \emph{point-set embedding} $\Gamma$ of $G$ on $S$ is a drawing of $G$ such that the vertices are placed to the points of $S$ through a one-to-one mapping, which can be given as part of the input (see, e.g.~\cite{DBLP:journals/gc/PachW01}) or not (see, e.g.~\cite{DBLP:journals/jgaa/KaufmannW02}). Fink et al. studied point-set RAC embeddings of graphs, where the points of $S$ belong to an $n \times n$ grid and no two points are horizontally or vertically aligned~\cite{DBLP:conf/walcom/FinkHMSW12}. They concentrate on computing point-set RAC embeddings with few bends per edge, where bends must occupy grid points as for the vertices. Different algorithmic results and bounds on the number of bends per edge are given, also under the assumption that the edge segments are restricted to be on grid lines. Among their results, they prove that every graph with $n$ vertices and $m$ edges admits a point-set RAC embedding on any $n \times n$ grid point set with at most three bends per edge in $O((n+m)^2)$ area. Note that, this result implies that every planar graph $G$ admits a RAC drawing with at most three bends per edge in quadratic area, which improves a previous result~\cite{DBLP:conf/s-egc/AngeliniBDFHKLL11} in terms of maximum number of bends per edge (see also Section~\ref{ss:tradeoffs-planar}).
On the negative side, they show that testing whether a graph admits a point-set RAC embedding without bends is \nph. 

%\subsection{Open Problems}
\myparagraph{Open Problems.} Although the study of  planar book embeddings has a long tradition in graph drawing (see, e.g.,~\cite{Ollmann-73,bk-79,DBLP:journals/jcss/Yannakakis89}), the study of its beyond-planar counterpart is much more recent. The question can be studied by either considering $k$-page drawings with forbidden crossing configurations (see, e.g.,~\cite{DBLP:journals/jgaa/BinucciCDGKKMT17}) or by focusing on $k$-page book embeddings of beyond-planar graphs (see, e.g.,~\cite{DBLP:journals/algorithmica/BekosBKR17,DBLP:journals/corr/AlamBK15,JGAA-454}). For example, we mention the following two problems.

\begin{problem}\label{op:2-page-dr}
Is there a function $f(\cdot)$ such that every planar graph of vertex-degree at most $\Delta$ admits a 2-page drawing that is $f(\Delta)$-planar?
\end{problem}

\begin{problem}\label{op:k-pl-bt}
Establish tight bounds on the book thickness of $k$-planar graphs. This question is interesting also when $k=1$.
\end{problem}

Every pair of planar graphs sharing their vertex set admits a simultaneous RAC embedding with at most six bends per edge~\cite{DBLP:journals/jgaa/BekosDKW16}. On the other hand, if no restriction on the edge crossings is given, such a pair admits a simultaneous embedding with at most two bends per edge~\cite{DBLP:journals/ijcga/GiacomoL07}.

\begin{problem}\label{op:sim-RAC}
Does every pair of planar graphs that share their vertex set admit a simultaneous RAC embedding with less than six bends per edge?
\end{problem}

A set $S$ of points in the plane is {\em universal} for  a family of graphs  if every element of the family admits a point-set embedding on $S$ with straight-line edges. A well-known result is that a universal point set of size $n$ that supports straight-line crossing-free drawings of all planar graphs with $n$ vertices does not exist~\cite{DBLP:journals/combinatorica/FraysseixPP90}. One may wonder whether a beyond-planar version of this problem is more likely to have a positive answer. In particular, we ask the following.

\begin{problem}\label{op:universal-k-planar}
Is there a set $S$ of $n$ points in the plane such that every planar graph with $n$ vertices admits an $f(n)$-planar point-set embedding on $S$ with straight-line edges such that $f(n) \in o(n)$?
\end{problem}

\section{Experiments and Engineering}\label{se:experiments-engineering}
The applied research in graph drawing beyond planarity has been mainly pursued along two directions: $(i)$ Cognitive experiments aimed to better understand the impact of some types of forbidden configurations on the capability of a user to execute visualization-based analysis tasks. $(ii)$ The implementation and experimentation of algorithms
to compute beyond-planar graph drawings, and the development of visualization systems for real-world application domains. We briefly discuss the contributions in these two directions.

\myparagraph{Cognitive studies.} Early cognitive experiments on beyond-planar graphs estimate the effects of crossing angles on human eye movements and performance, through the use of eye-tracking systems~\cite{DBLP:conf/apvis/Huang07,DBLP:journals/puc/Huang13}. The results show that sharp angles may trigger extra eye movements, causing delays for path search tasks, whereas crossings have usually little impact on node locating tasks.
Subsequent experiments confirm the importance of large angle crossings to execute visualization-based analysis tasks~\cite{DBLP:conf/apvis/HuangHE08,DBLP:journals/vlc/HuangEH14}, thus motivating and stimulating the rich literature about RAC graphs and related graph families~\cite{dl-13}. Further studies give some evidence that relevant improvements on the readability of graph layouts derive from finding a good trade-off between different quality metrics, rather than optimizing only one of them; the considered metrics include both crossing angles and vertex angles~\cite{DBLP:journals/vlc/HuangEHL13}.

We recall that Partial Edge Drawing (PED) is a graph drawing style aimed at reducing edge crossings and visual clutter. PEDs are straight-line drawings where the central part of each edge is erased, and the length of the two remaining segments are computed so to preserve useful geometric information~\cite{bew-vnd-TVCG95}.
Given a straight-line drawing, an $\alpha$-SHPED  of this drawing is immediately defined for a fixed value of $\alpha$~\cite{DBLP:conf/fun/BruckdorferK12}. In particular, some edge crossings may not be avoidable, although the amount of ink removed from the original drawing might be large (e.g., $50\%$ when $\alpha=\frac{1}{4}$). On the other hand, it is possible to maximize the ink and remove edge crossings by renouncing to homogeneity. Binucci et al. present a user study in which PEDs obtained via heuristics are compared with the standard model $\frac{1}{4}$-SHPED~\cite{DBLP:conf/iisa/BinucciLMT16}. The results suggest that the benefit of homogeneity overcomes in terms of readability the benefit of fewer crossings and more ink.

Another line of user experiments on beyond-planar graphs focuses on assessing a so-called \emph{edge stratification approach} to analyze complex visualizations of graphs~\cite{DBLP:journals/vlc/GiacomoDLMT14}. The approach is based on partitioning the edge set of a drawing into a minimal number of layers, such that the edges in each layer define a drawing with some desired properties related to crossings. Other than requiring that the drawing of each layer is planar, other possibilities are that the drawing of each layer has crossing angles larger than a given constant $\alpha$, or that it is $k$-planar for some fixed $k$. The experiments show that the stratification approach is mainly useful for local tasks such as counting the degree of a vertex, while it is less effective for more global tasks such as finding shortest paths between pairs of vertices. The edge stratification algorithms proposed by Di Giacomo et al. are sometimes computationally expensive, especially when the drawing on each layer is required to be $k$-planar for relatively large values of $k$; most of these algorithms can be successfully applied to drawings with few hundred vertices and edges~\cite{DBLP:journals/vlc/GiacomoDLMT14}.

\begin{figure}[tb]
	\centering
	\subfigure[]{\includegraphics[width=0.24\columnwidth]{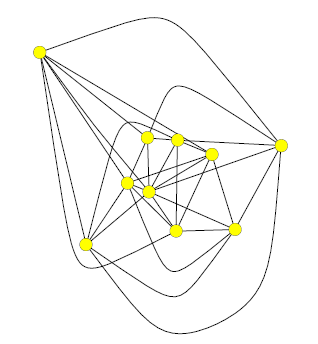}\label{fi:tdfd-b}}\hfill
	\subfigure[]{\includegraphics[width=0.24\columnwidth]{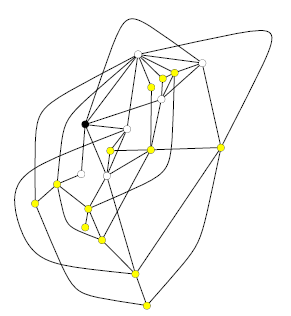}\label{fi:tdfd-a}}\hfill
	\subfigure[]{\includegraphics[width=0.5\columnwidth]{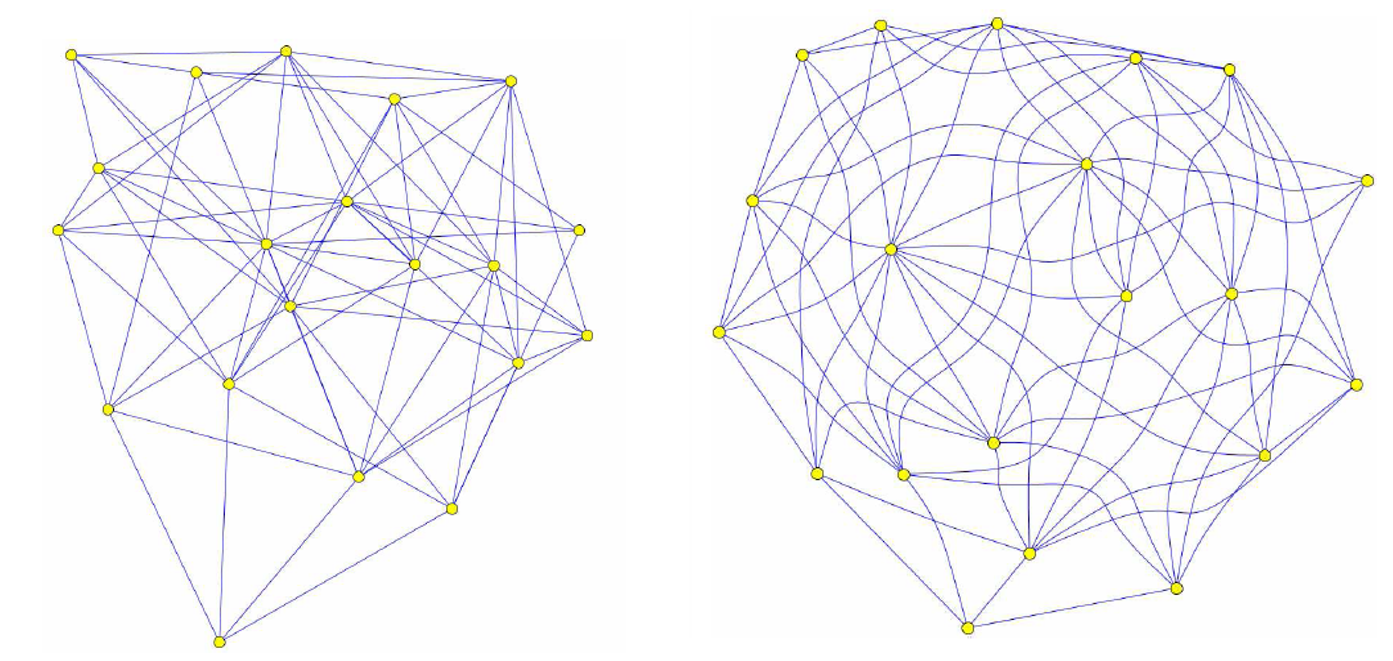}\label{fi:tdfd-d}}
	%\subfigure[]{\includegraphics[width=0.24\columnwidth]{figs/tdfd-c}\label{fi:tdfd-c}}
	\caption{(a)-(b) Examples of drawings computed with the topology-driven force-directed approach. (c) A straight-line drawing computed with a force-directed algorithm (left) and a subsequent drawing with improved crossing angle resolution, obtained using smoothed curves (right).}\label{fi:tdfd}
\end{figure}

\myparagraph{Algorithms and systems.} Several papers describe heuristics that attempt to optimize some desired properties for edge crossings, often in combination with classical graph drawing conventions. For example, in order to improve the readability of circular layouts, a post-processing algorithm, called \textsf{MAXCIR}~\cite{DBLP:conf/gd/NguyenEHH10}, aims to increase crossing angles by using Quadratic Programming; experiments show that this algorithm is fast in practice and yields better results compared to a traditional equal-spacing algorithm. \textsf{BIGANGLE}~\cite{DBLP:journals/vlc/HuangEHL13} is a force-directed algorithm that computes drawings of graphs with multiple aesthetics being improved at the same time; these aesthetics include crossing and vertex angle resolution. Argyriou et al. describe the design and the implementation of another force-directed algorithm that computes drawings with low crossing and vertex angle resolution~\cite{DBLP:journals/cj/ArgyriouBS13}.
A further contribution in the same direction as \textsf{BIGANGLE} is the \emph{topology-driven force-directed framework}~\cite{DBLP:conf/gd/DidimoLR10}. It allows the design of graph drawing algorithms that find good trade-offs between different aesthetics such as number of crossings, crossing angle resolution, geodesic edge tendency, and number of edge bends. This approach is based on combining force-directed techniques with the popular topology-shape-metrics approach, originally proposed for the computation of bend-minimum orthogonal drawings of graphs~\cite{DBLP:journals/siamcomp/Tamassia87}.
Examples of drawings computed by topology-driven force-directed algorithms are shown in Figs.~\ref{fi:tdfd-b} and~\ref{fi:tdfd-a}. As a follow-up of this work, the same authors implement an algorithm that first computes a straight-line drawing with a force-directed technique and then it runs a post-processing procedure to improve crossing angle resolution by representing edges as smoothed curves (see, e.g., Fig.~\ref{fi:tdfd-d}); each curve is monotone in the direction of the straight-line segment connecting its end-points. The algorithm is also applied to the simultaneous embedding of suitably defined networks in a system for conceptual Web-site traffic analysis~\cite{DBLP:journals/jgaa/DidimoLR11}.
%, called \textsf{COWA}~\cite{DBLP:journals/jgaa/DidimoLR11}; see, e.g., Fig.~\ref{fi:cowa}).

With the aim of computing layered drawings of graphs with large-angle crossings (see Section~\ref{se:constraints}), Di Giacomo et al. describe building block heuristics for extracting a maximum 2-layer RAC subgraph of a given bipartite graph~\cite{DBLP:journals/cj/GiacomoDGLR15}. They study both the setting in which the input graph has no fixed ordering for the vertices of each partition set and the setting in which one of the two sets has a given linear order, which must be preserved by the RAC subgraph.

We finally mention the implementation of an algorithm that computes ortho-polygon visibility representations (OPVRs) of (non-planar) graphs~\cite{DiGiacomo2017}. As explained in Section~\ref{ss:Vertex-Coplexity}, in an OPVR of a graph $G$, each vertex is represented as an orthogonal polygon and each edge is either a horizontal or a vertical segment. %(see, e.g., Fig.~\ref{fi:orthovisi}). 
The \emph{vertex complexity} of an OPVR is the maximum number of reflex angles inside a polygon representing a vertex. 
%For example, the OPVR in Fig.~\ref{fi:orthovisi} has vertex complexity three. 
Assuming that the graph $G$ comes with a fixed embedding, the algorithm in~\cite{DiGiacomo2017} tests in $O(n^2)$ time whether an OPVR of $G$ exists and, if so, it computes in $O(n^\frac{5}{2}\log^\frac{3}{2}n)$ time an embedding-preserving OPVR of $G$ with minimum vertex complexity. This algorithm has also been experimented on a large set of $1$-plane graphs ($1$-plane graphs always admit an OPVR). The experimental results show that, in practice, the computed OPVRs have usually a high percentage (up to $90\%$ in some cases) of vertices drawn with no reflex corners (i.e., as rectangles). 
%(see the darker vertices in the representation of Fig.~\ref{fi:orthovisi}).

%\begin{figure}[tb]
%	\centering
%	\subfigure[]{\includegraphics[width=0.42\columnwidth]{figs/cowa}\label{fi:cowa}}\hfill
%	\subfigure[]{\includegraphics[width=0.42\columnwidth]{figs/orthovisi}\label{fi:orthovisi}}
%	\caption{(a) Simultaneous embedding of two conceptual networks in \textsf{COWA}. (b) OPVR of a $1$-plane graph.}\label{fi:sys}
%\end{figure}

%\subsection{Open Problems}
\myparagraph{Open Problems.} There is a general lack of user-studies to assess the effectiveness of the beyond-planar graph models in practice. For example, a variant of PEDs for orthogonal drawings has been recently introduced~\cite{DBLP:journals/jgaa/BruckdorferKM14} but its effectiveness in practice has not yet been investigated.

\begin{problem}\label{op:orto-ped}
Perform a user study in order to understand whether orthogonal PEDs are more readable than traditional orthogonal drawings.
\end{problem}

One of the most popular techniques to draw graphs is the so-called ``Sugiyama's approach''~\cite{DBLP:journals/tsmc/SugiyamaTT81}. We recall here that this approach places the vertices of the graph on several layers (horizontal lines) and one of its main objectives is to minimize the edge crossings between two consecutive layers (see~\cite{dett-gd-99} for a comprehensive description of this approach).  As mentioned in the introduction, Mutzel observed that the readability of a $2$-layer drawing may not just depend on the number but also on the type of edge crossings~\cite{DBLP:journals/siamjo/Mutzel01}. Hence we propose the study of the following problem, for which some efforts have been limited so far to RAC drawings~\cite{DBLP:journals/cj/GiacomoDGLR15}. 

%it includes a crucial step called crossing minimization. In this step the input is a bipartite graph  $G(V_1,V_2,E)$ and a permutation $\pi_1$ of $V_1$;  the output is a $2$-layer drawing of $G$ such that the vertices $V_1$ respect permutation $\pi_1$ on a layer while the linear order of the vertices of $V_2$ is chosen so to reduce the number of edge crossings (see also Section~\ref{se:constraints} for a definition of $2$-layer drawings). As recalled in the introduction, Mutzel~\cite{DBLP:journals/siamjo/Mutzel01} observed that the readability of a $2$-layer drawing may not just depend on the number but also on the type of edge crossings. Hence we propose the following open problem, where by ``traditional crossing minimization heuristics'' we mean those that are more commonly adopted in the Sugiyama's approach (such as, for example, the median heuristic, the barycenter heuristic, the split heuristic etc.).

%\begin{problem}\label{op:sugyiama}
%Design heuristics that reduce forbidden crossing configurations in a 2-layered drawing of a bipartite graph $G(V_1,V_2,E)$ where the linear order for the vertices of $V_1$ is fixed while the vertices of $V_2$ can be arbitrarily permuted. Integrate these heuristics in the Sugiyama's approach and perform a user-study to compare the obtained drawings with those computed by means of the traditional crossing minimization heuristics.
%\end{problem}

\begin{problem}\label{op:sugiyama}
Design heuristics that reduce the number of forbidden crossing configurations in $2$-layered drawings of bipartite graphs, integrate them into Sugiyama's approach, and perform a user-study to compare the obtained drawings with those computed by means of $2$-layer crossing minimization heuristics.
\end{problem}

Finally, it would be interesting to analyze the practical impact that the various forbidden edge crossing configurations have in real-world drawings. For example, one could consider a large benchmark of real-world graphs with various sizes, draw these graphs by using different algorithms (e.g., force-directed algorithms) and then analyze the frequency of occurrence of each forbidden edge crossing configuration in these drawings. We summarize this problem as follows.

\begin{problem}\label{op:forb-conf}
Analyze the impact of various forbidden edge crossing configurations in a large set of real-world drawings.
\end{problem}

%
%\section{Beyond this Survey}\label{se:related}
%\input{beyond}
%
\section{Concluding Remarks}\label{se:conclusions}
The treatment of edge crossings in graph visualization embraces other topics that have not been discussed in the previous sections. Although these topics are not properly recognized as part of the literature on graph drawing beyond planarity, recent works have established interesting connections with beyond-planar graph families, which highlight new interesting research directions.

\myparagraph{Crossing Minimization.} Computing drawings of graphs with minimum number of crossings is a problem with a long tradition, and a huge amount of papers has been devoted to it (see~\cite{DBLP:reference/crc/BuchheimCGJM13} for a survey). It is well known that the crossing minimization problem is \npc and it remains hard also in very restricted scenarios, for instance when the graph is bipartite and we look for a straight-line $2$-layer drawing of it~\cite{gj-83}. However, for $2$-layer RAC drawable graphs the problem can be solved efficiently~\cite{DBLP:journals/algorithmica/GiacomoDEL14}, which suggests that the crossing minimization problem might be polynomial-time solvable also with respect to other forbidden crossing configurations.

\myparagraph{Relaxed Clustered Planarity.} In a \emph{clustered graph}, vertices are grouped into (hierarchical) clusters. In a drawing of a clustered graph each cluster should be clearly represented as a closed region containing all (and only) its vertices and all (and only) its subclusters. Also, an edge should not traverse (i.e., cut) a cluster if both its endvertices are outside the cluster. Many papers study how to efficiently test whether a clustered planar graph admits a crossing-free drawing that meets the aforementioned properties~\cite{kw-dg-01,DBLP:conf/compgeom/CorteseB05}, and practical drawing heuristics have also been conceived, even for non-planar and large graphs (e.g.,~\cite{DBLP:conf/gd/BattistaDM01,DBLP:conf/iv/BourquiAM07,DBLP:journals/isci/DogrusozGCCD09,DBLP:journals/isci/DidimoM14}). In the spirit of graph drawing beyond planarity, relaxed models of clustered planarity can be studied, in which only some types of crossings are disallowed. Initial steps in this direction can be found in~\cite{DBLP:journals/comgeo/AngeliniLBFPR15}.

\myparagraph{Hybrid Visualizations.} The use of matrix-based representations combined with the classical node-link representation has been proposed to diminish the negative effect of visual clutter in large and locally dense networks~\cite{DBLP:journals/tvcg/HenryFM07,DBLP:journals/tvcg/BatageljBDLPP11}. More recently, the planarity testing and embedding problem of this type of hybrid visualizations have been further formalized~\cite{DBLP:journals/jgaa/LozzoBFP18,nodetrix-gd17}, and new families of beyond-planar graphs related to hybrid visualizations have been defined~\cite{planarknodetrix-gd17}.

\myparagraph{Edge Bundling.} The idea of grouping edges to get planar drawings of non-planar graphs was originally proposed under the name of \emph{confluent drawings}~\cite{DBLP:journals/jgaa/DickersonEGM05,DBLP:journals/algorithmica/EppsteinGM07}. Later on, practical edge-bundling techniques were used to cope with the visual clutter problem in node-link diagrams of large and dense graphs~\cite{DBLP:journals/tvcg/Holten06,DBLP:journals/cgf/HoltenW09,zxyq-13}. The idea is that ``similar'' edges are deformed and grouped into bundles, thus providing a more abstract and uncluttered view of the original drawing at the expenses of possible ambiguities in terms of connections between vertices. Very recently, the use of edge bundling has been proposed in the context of graph drawing beyond planarity, with the introduction of the family of \emph{1-fan-bundle planar} graphs, which combines edge bundling with fan-planar graphs~\cite{1-fbp-gd17}. Following this example, other families can be conceived and studied.

%\smallskip Finally, we remark that the seminal intuition by Mutzel about the topology of the crossings in a drawing started from an empirical observation. Also the definition of RAC drawings is the consequence of cognitive studies on how people read diagrams. In order to develop effective systems that are useful to compute drawings with "readable" crossings, the algorithmic community should perhaps take into account that the optimization goal about what are the acceptable crossing configurations must be defined by first executing cognitive studies on how people read diagrams and then, based on such experiments define new families of beyond-planar graphs and develop theories and algorithms about them.

\smallskip Finally, we remark that a very promising research direction is to perform new experiments and develop new theories on graph drawing beyond planarity based on the following user-centered approach:  (i) Develop theories on how people read graphs with crossings based on HCI experiments; (ii) Define Optimization criteria on the edge crossings and their configurations; (iii) Design  combinatorial models and efficient algorithms; (iv) Experimentally verify the efficiency and effectiveness of the algorithms with new experiments that may lead to refining the model and/or the optimization goals.

\bibliographystyle{abbrv}
\bibliography{beyondplanarity}

\end{document}